\documentclass[twocolumn,useamsfonts]{pasj01}

\usepackage{graphicx}
\usepackage{dcolumn}
\usepackage{amssymb} 
\usepackage{ctable}
\usepackage{bm}
\usepackage{rotating}
\usepackage{natbib}
\usepackage{array}
\usepackage{aas_macros}
\usepackage[T1]{fontenc}
\usepackage{lmodern}

\newcommand{\cmnt}[1]{}

\newcommand{\oiii}{[O\,\emissiontype{III}]}
\newcommand{\oii}{[O\,\emissiontype{II}]}
\newcommand{\nii}{[N\,\emissiontype{II}]}
\newcommand{\sii}{[S\,\emissiontype{II}]}

\newenvironment{fund}{\section*{Funding}\fontsize{8}{11}\selectfont}{\par}

\def\vecp{\mathbf{p}}

\def\vecd{\mathbf{d}}
\def\estd{\mathbf{\hat{d}}}
\def\matf{\mathbf{F}}
\def\vecq{\mathbf{q}}
\def\vecx{\mathbf{x}}
\def\vecy{\mathbf{y}}
\def\matc{\mathbf{C}}

\def\veck{\mathbf{k}}

\def\VEV#1{\left\langle #1 \right\rangle}

\def\ovn{\overline{n}}

\begin{document}
\bibliographystyle{apj}

\title{Interloper bias in future large-scale structure surveys}

\author{Anthony R. \textsc{Pullen}\altaffilmark{1,2,3}, Christopher M. \textsc{Hirata}\altaffilmark{4,3}, Olivier \textsc{Dor\'{e}}\altaffilmark{2,3}, and Alvise \textsc{Raccanelli}\altaffilmark{5,2,3}}
\altaffiltext{1}{Department of Physics, Carnegie Mellon University, 5000 Forbes Ave, Pittsburgh, Pennsylvania, 15213, U.S.A.}
\altaffiltext{2}{NASA Jet Propulsion Laboratory, California Institute of Technology, 4800 Oak Grove Drive, Pasadena, CA, 91109, U.S.A.}
\altaffiltext{3}{California Institute of Technology, Pasadena, California, 91125 U.S.A.}
\altaffiltext{4}{Center for Cosmology and Astroparticle Physics, The Ohio State University, 191 West Woodruff Avenue, Columbus, Ohio, 43210, U.S.A.}
\altaffiltext{5}{Department of Physics and Astronomy, John Hopkins University, 3400 N. Charles St,
Baltimore, Maryland, 21218, U.S.A.}
\email{apullen@andrew.cmu.edu}

\Received{\today}

\maketitle


\begin{abstract}
Next-generation spectroscopic surveys will map the large-scale structure of the observable universe, using emission line galaxies as tracers.  While each survey will map the sky with a specific emission line, interloping emission lines can masquerade as the survey's intended emission line at different redshifts.  Interloping lines from galaxies that are not removed can contaminate the power spectrum measurement, mixing correlations from various redshifts and diluting the true signal.  We assess the potential for power spectrum contamination, finding that an interloper fraction worse than 0.2\% could bias power spectrum measurements for future surveys by more than 10\% of statistical errors, while also biasing power spectrum inferences.  We also construct a formalism for predicting cosmological parameter measurement bias, demonstrating that a 0.15--0.3\% interloper fraction could bias the growth rate by more than 10\% of the error, which can affect constraints on gravity from upcoming surveys.  We use the COSMOS Mock Catalog (CMC), with the emission lines re-scaled to better reproduce recent data, to predict potential interloper fractions for the Prime Focus Spectrograph (PFS) and the Wide-Field InfraRed Survey Telescope (WFIRST).  We find that secondary line identification, or confirming galaxy redshifts by finding correlated emission lines, can remove interlopers in PFS. For WFIRST, we use the CMC to predict that the 0.2\% target can be reached for the WFIRST H$\alpha$ survey, but sensitive optical and near-infrared photometry will be required. For the WFIRST \oiii\ survey, the predicted interloper fractions reach several percent and their effects will have to be estimated and removed statistically (e.g. with deep training samples). These results are optimistic as the CMC does not capture the full set of correlations of galaxy properties in the real Universe, and they do not include blending effects. Mitigating interloper contamination will be crucial to the next generation of emission line surveys. 
\end{abstract}

\section{Introduction}
A new generation of large-scale structure (LSS) surveys will come online in the next decade.  In particular, there is much activity on the spectroscopic front, as many ground-based surveys, including the Prime Focus Spectrograph (PFS) \citep{2014PASJ...66R...1T}, FastSound \citep{2015PASJ...67...81T}, the Hobby-Eberly Dark Energy eXperiment (HETDEX) \citep{2004AIPC..743..224H,2008ASPC..399..115H}, and the Dark Energy Spectroscopic Instrument (DESI) \citep{2013arXiv1308.0847L}, expect to take data in the next several years, while the space-based surveys \emph{Euclid} \citep{2011arXiv1110.3193L} and the Wide-Field Infrared Survey Telescope (WFIRST) \citep{2013arXiv1305.5422S} plan to come online soon after. As more surveys are going deeper and wider to find more galaxies and collect better statistics on LSS, it is becoming more important to identify systematic effects that can contaminate cosmological parameter estimates. Moreover, in order to observe more galaxies at higher redshift, these future surveys will acquire spectra with lower signal-to-noise ratio than were obtained by past projects such as the Sloan Digital Sky Survey (SDSS) \citep{2000AJ....120.1579Y}, which focused on relatively bright targets. SDSS galaxy spectra usually had many spectral features readily identifiable by eye, but future LSS surveys are likely to operate near the minimum acceptable signal-to-noise ratio for redshift determination using a \emph{single} emission line.

One systematic effect that could potentially contaminate future LSS measurements is interloper contamination.  The Hubble expansion of the universe causes extragalactic emission lines to appear redshifted.  While the redshift of the emission line can identify the object's Hubble velocity and distance, this effect can cause two lines with different rest-frame wavelengths from different distances to appear with the same wavelength, confusing the two lines.  Interlopers can be problematic in power spectrum estimates as well as measurements of the correlation function, baryon acoustic oscillations (BAO), weak lensing, redshift-space distortions (RSD), scale-dependent clustering bias, and higher-order correlations such as the bispectrum because interlopers introduce an extra source of correlation and dilute the existing correlation, distorting the power spectrum.  These distortions can contaminate constraints on cosmological parameters, including those for the growth rate of structure.  Interlopers are also a major consideration in intensity mapping surveys \citep{2010JCAP...11..016V,2014ApJ...786..111P,2014ApJ...785...72G}, where the data product is a low-resolution data cube (RA,Dec,$\lambda$) and the problem {\em must} be treated statistically instead of via object-by-object redshift classification. While it is possible to remove this distortion by measuring the interloper fraction in a small patch of sky, the required precision may be formidable in cases where the power spectrum measurement is very sensitive to the interloper fraction.

Methods have been developed to identify interloping emission lines, including simple methods such as secondary line identification and photometry \citep{2007ApJ...660...62K} as well as more complex methods such as spectral template fitting.  However, these methods are limited by the LSS survey parameters, including photometric bands, spectral coverage, survey depth, and (for ground surveys) atmospheric lines; thus, it is important to diagnose for a specific LSS survey how well interlopers can be identified within the data.

In this paper, we construct a formalism describing how a power spectrum from a LSS survey is distorted by interloper contamination.  This formalism is then used to determine the resulting bias within measurements of cosmological parameters.  We then calculate the interloper bias for the clustering bias of sample $b_g$ and the growth rate $f_g$  as a function of the interloper fraction.  We find that for PFS and WFIRST, an interloper fraction $\gtrsim 0.2$\% will significantly distort power spectrum measurements over scales $k>0.01h$Mpc$^{-1}$ such that the $k$-averaged power spectrum amplitude will shift by $\gtrsim10$\% of the amplitude error. We also find that an interloper fraction greater than 0.3 (0.15)\% can significantly bias growth rate measurements for PFS (WFIRST).

To test if these biases will be exhibited in future surveys, we use the COSMOS \citep{2009ApJ...690.1236I} Mock Catalog (CMC) \citep{2011A&A...532A..25J} to construct mock surveys for the PFS \oii\ survey and the WFIRST H$\alpha$ and \oiii\ surveys based on their flux sensitivity curves, for the cases of no spectral cleaning and with cleaning using secondary line identification.  Note that our analysis with the CMC is idealized in that it does not treat blended objects and is limited by the bank of SED templates in the catalog.  We also made modifications in the CMC to better reproduce the properties of real objects in the Universe. But it does represent a ``first look'' at how serious the problem might be.  Subject to these caveats, we determine that after secondary line identification, PFS will have an interloper fraction less than 0.2\%, meaning interlopers may not be a great concern.  For WFIRST, on the other hand, secondary line identification will only reduce some interloper fractions to 10--30\%.  We then determine photometric cuts using infrared photometric bands from WFIRST and optical bands from the Large Synoptic Survey Telescope (LSST) \citep{2012arXiv1211.0310L} that reduce the interloper fractions to less than 0.2\% at most redshifts in the H$\alpha$ survey.  In the WFIRST \oiii\ survey, the interloper fractions in WFIRST are still very large, and deep spectroscopic fields will be needed to precisely measure the interloper fractions.  Of course, deep surveys will also be needed for the PFS and WFIRST H$\alpha$ surveys to confirm if interloper fractions for these surveys are indeed small.  We also show that the WFIRST Y band depth will be necessary for interloper removal.  Future LSS emission line surveys must consider how to implement these strategies to identify interlopers, as well as determine acceptable contamination levels for other cosmological parameters.

The plan of our paper is as follows: in Sec.~\ref{S:bias}, we derive the interloper bias to the measured galaxy power spectrum and cosmological parameters as functions of the interloper fraction and redshift.  We present methods to identify and remove interlopers in Sec.~\ref{S:removal}.  In Sec.~\ref{S:mock}, we describe the mock surveys we construct for PFS and WFIRST using the CMC, and we give potential interloper fractions for PFS in Sec.~\ref{S:pfsresults} and WFIRST in Sec.~\ref{S:wfresults}.  We state our conclusions in Sec.~\ref{S:conc}.  Wherever not explicitly mentioned, we assume a flat $\Lambda$CDM cosmology with parameters compatible with the Wilkinson Microwave Anisotropy Probe Seven-Year Data Release (WMAP7) \citep{2011ApJS..192...14J}.

\section{Interloper bias} \label{S:bias}
In this section we derive the effect of interlopers on the matter power spectrum measurement, as well as the bias introduced in cosmological parameter measurements.  We consider a hypothetical survey that maps survey emission line galaxies (SELGs), detected through a survey emission line (SEL) with rest-frame wavelength $\lambda_{\rm SEL}$, and is contaminated by an interloping emission line with rest-frame wavelength $\lambda_{\rm Int}$.  Note in this section we assume only one interloper, but we can easily extend our analysis to multiple interlopers.  Since emission line $\lambda_{\rm SEL}$ from redshift $z_{\rm SELG}$ and emission line $\lambda_{\rm Int}$ at redshift $z_{\rm Int}$ are observed at the same wavelength, each $z_{\rm SELG}$ has a corresponding $z_{\rm Int}$ according to
\begin{eqnarray}\label{E:zint}
1+z_{\rm Int}=\frac{\lambda_{\rm SEL}}{\lambda_{\rm Int}}(1+z_{\rm SELG})\, .
\end{eqnarray}

\subsection{Galaxy Power Spectrum}
We consider a galaxy power spectrum measurement constructed from overdensity maps of number counts of galaxies, where each galaxy emits an emission line that appears to be a SEL with wavelength $\lambda=\lambda_{\rm SEL}(1+z_{\rm SELG})$.  However, the number counts will include not only SELGs but interloping galaxies as well.  Thus, the overdensity is written in terms of the total \emph{comoving} number density $n_t=n_{\rm SELG}+n_{\rm Int}$ in the form
\begin{eqnarray}
\delta_t(\vecx)&=&\frac{n_t(\vecx)-\ovn_t}{\ovn_t}\nonumber\\
&=&\frac{n_{\rm SELG}(\vecx)-\ovn_{\rm SELG}}{\ovn_t}\nonumber\\
&&+\frac{n_{\rm Int}(\vecy)-\ovn_{\rm Int}}{\ovn_t}\, ,
\end{eqnarray}
where $n_{\rm Int}$ is the number of interlopers divided by the SELG volume element (not redshifted to $z_{\rm Int}$), the barred quantities ($\ovn$) are averaged over the survey area, and $\vecx$ and $\vecy$ are the 3D comoving position vectors of SELGs and interloping galaxies, respectively.  $\vecy$ is the true position of an interloping galaxy at redshift $z_{\rm Int}$ such that it appears to be at position $\vecx$ if it is assumed to be at redshift $z_{\rm SELG}$.  We can separate the position vectors in terms of their transverse ($\vecx_\perp$) and radial ($\vecx_\parallel$) components, and it can be shown that $\vecx_\perp\propto D(z_{\rm SELG})$, the comoving distance to redshift $z_{\rm SELG}$, and $\vecx_\parallel\propto(1+z_{\rm SELG})/H(z_{\rm SELG})$ (similar expressions are true for $\vecy$).  Thus, with $\vecx$ being the observed position of the interloper and $\vecy$ being its true position, we define $\gamma_\perp$ and $\gamma_\parallel$ such that $(\vecx_\perp,\vecx_\parallel)=(\gamma_\perp \vecy_\perp,\gamma_\parallel \vecy_\parallel)$ with
\begin{eqnarray} \label{E:gamma}
\gamma_\perp&=&\frac{D(z_{\rm SELG})}{D(z_{\rm Int})}\nonumber\\
\gamma_\parallel&=&\frac{(1+z_{\rm SELG})/H(z_{\rm SELG})}{(1+z_{\rm Int})/H(z_{\rm Int})}\nonumber\\
&=&\frac{\lambda_{\rm Int}H(z_{\rm Int})}{\lambda_{\rm SEL}H(z_{\rm SELG})}\, .
\end{eqnarray}
It is evident from Eqs.~\ref{E:zint} and \ref{E:gamma} that $\gamma_\perp>1$ ($\gamma_\perp<1$) for $\lambda_{\rm Int}>\lambda_{\rm SEL}$ ($\lambda_{\rm Int}<\lambda_{\rm SEL}$).  For $\gamma_\parallel$, $\lambda_{\rm Int}>\lambda_{\rm SEL}$ causes $H(z_{\rm Int})/H(z_{\rm SELG})<1$, and vice-versa, making the behavior of $\gamma_\parallel$ less trivial.  For $\lambda_{\rm Int}>\lambda_{\rm SEL}$, we find $\gamma_\parallel>1$ ($\gamma_\parallel<1$) for $z_{\rm SEL}<z_\Lambda$ ($z_{\rm SELG}>z_\Lambda$) where $\Lambda=\lambda_{\rm Int}/\lambda_{\rm SEL}$ and
\begin{eqnarray}
1+z_\Lambda=\sqrt[3]{\frac{1-\Omega_m}{\Omega_m}\Lambda(\Lambda+1)}\, ,
\end{eqnarray}
for a flat $\Lambda$CDM universe.  The opposite is true for $\lambda_{\rm Int}<\lambda_{\rm SEL}$.

We now assert the existence of an interloper fraction $f$, such that the number density of interlopers is $f$ times the total number density of objects, or $\ovn_{\rm Int}=f\ovn_t$.  Thus, $f=0$ is a map with no interloper contamination, and $f$ approaches unity as the interloper contamination increases.  The quantity $f$ is averaged over the sky and over each redshift bin, so that it does not contain LSS fluctuations and is not a random field.  We write $\ovn_t$ in terms of $f$, $\ovn_{\rm SELG}$, and $\ovn_{\rm Int}$, according to
\begin{eqnarray}\label{E:fn}
\ovn_t=\frac{\ovn_{\rm SELG}}{1-f}=\frac{\ovn_{\rm Int}}{f}\, .
\end{eqnarray}
Thus, we can separate $\delta_t$ into the overdensities of the two sets of galaxies, in the form
\begin{eqnarray}
\delta_t(\vecx)=(1-f)\delta_{\rm SELG}(\vecx)+f\delta_{\rm Int}(\vecy)\, .
\end{eqnarray}
A similar equation exists for the Fourier transform of this expression, except that the Fourier transform of the interloper term becomes
\begin{eqnarray}
\delta_{\rm Int}^{\rm obs}(\veck)&=&\int d^3\vecx\,e^{i\veck\cdot\vecx}\delta_{\rm Int}(\vecy)\nonumber\\
&=&\gamma_\perp^2\gamma_\parallel\int d^3\vecy\,e^{i\vecq\cdot\vecy}\delta_{\rm Int}(\vecy)\nonumber\\
&=&\gamma_\perp^2\gamma_\parallel\delta_{\rm Int}(\vecq)\, ,
\end{eqnarray}
where $\vecq=(\veck_\perp\gamma_\perp,\veck_\parallel\gamma_\parallel)$.  This implies that the covariance of $\delta_{\rm Int}^{\rm obs}(\veck)$ is given by
\begin{eqnarray}
\VEV{\delta_{\rm Int}^{\rm obs}(\veck)\delta_{\rm Int}^{*{\rm obs}}(\veck')}\!\!\!\!
&=&\!\!\!\!
(\gamma_\perp^2\gamma_\parallel)^2P_{\rm Int}(\vecq)\delta_D(\vecq-\vecq')~~\nonumber\\
&=&\!\!\!\!
\gamma_\perp^2\gamma_\parallel P_{\rm Int}(\vecq)\delta_D(\veck-\veck')\, ,
\end{eqnarray}
where $\delta_D(\vecx)$ is a 3D delta function.

Since the two sets of galaxies are at very different redshifts, their overdensities should be uncorrelated.  In particular, of the cases considered in this paper, the most significant interloper effect is H$\alpha$ and \oiii\ ($\Delta\ln\lambda = 0.27$), implying that the redshift separation of the contaminants is $\Delta\ln(1+z)=0.27$.\footnote{An exception to this argument, where we would have to consider the correlation between the SEL and interloper density fields, would occur if $\lambda_{\rm SEL}\approx \lambda_{\rm Int}$. This does not occur in this paper, but it does occur in the case of Si {\sc iii} 1206 \AA\ contamination in the Lyman-$\alpha$ forest, where the target and contaminating lines are separated by only 2300 km/s; see e.g. \citet{2006ApJS..163...80M}.}  Note that gravitational lensing could induce small correlations \citep{2013arXiv1311.6813R,2015arXiv150506179R} between the SELGs and the interlopers, though we will neglect them in our analysis.
Thus, the total 3D galaxy power spectrum $P_t(k,\mu)$, including redshift-space distortions (RSD) with $\mu=k_\parallel/k$, can be written as the sum of the two individual components, according to\footnote{This expression was shown earlier in a private communication with D.~Eisenstein.}
\begin{eqnarray}
&& \!\!\!\!\!\!\!\!\!\!\!\!\!\!\!\!\!\!\!\!\!\!\!\!\!\!\!\!\! P_t(f|k,\mu,z_{\rm SELG}) 
\nonumber \\
&=& \!\!\!\! (1-f)^2P_{\rm SELG}(k,\mu,z_{\rm SELG})\nonumber\\
&& +f^2\gamma_\perp^2\gamma_\parallel P_{\rm Int}[q(k,\mu),\mu_q(\mu),z_{\rm Int}]\, ,
\end{eqnarray}
where $\gamma(\mu)=\sqrt{\gamma_\perp^2(1-\mu^2)+\gamma_\parallel^2\mu^2}$, $q(k,\mu)=\gamma(\mu)k$, $\mu_q(\mu)=\gamma_\parallel\mu/\gamma(\mu)$, and $P_{\rm SELG} (k,\mu,z_{\rm SELG})$ is given by \citep{1987MNRAS.227....1K, 1998ASSL..231..185H}
\begin{equation} \label{E:pkmu}
P_{\rm SELG} (k,\mu,z_{\rm SELG}) = (1+\beta\mu^2)^2 P_{g,r} \, ,
\end{equation}
where $P_{g,r}$ is the power spectrum in real space and $\beta = f_g/b_g$, with $b_g$ being the galaxy bias relating visible to dark matter halos.  Strictly speaking, the shot noise will also vary with the interloper fraction.  But since it is usually fitted and subtracted from the measured power spectrum, we only consider the interloper bias to the clustering power spectrum.

\begin{figure}
\begin{center}
\includegraphics[width=0.5\textwidth]{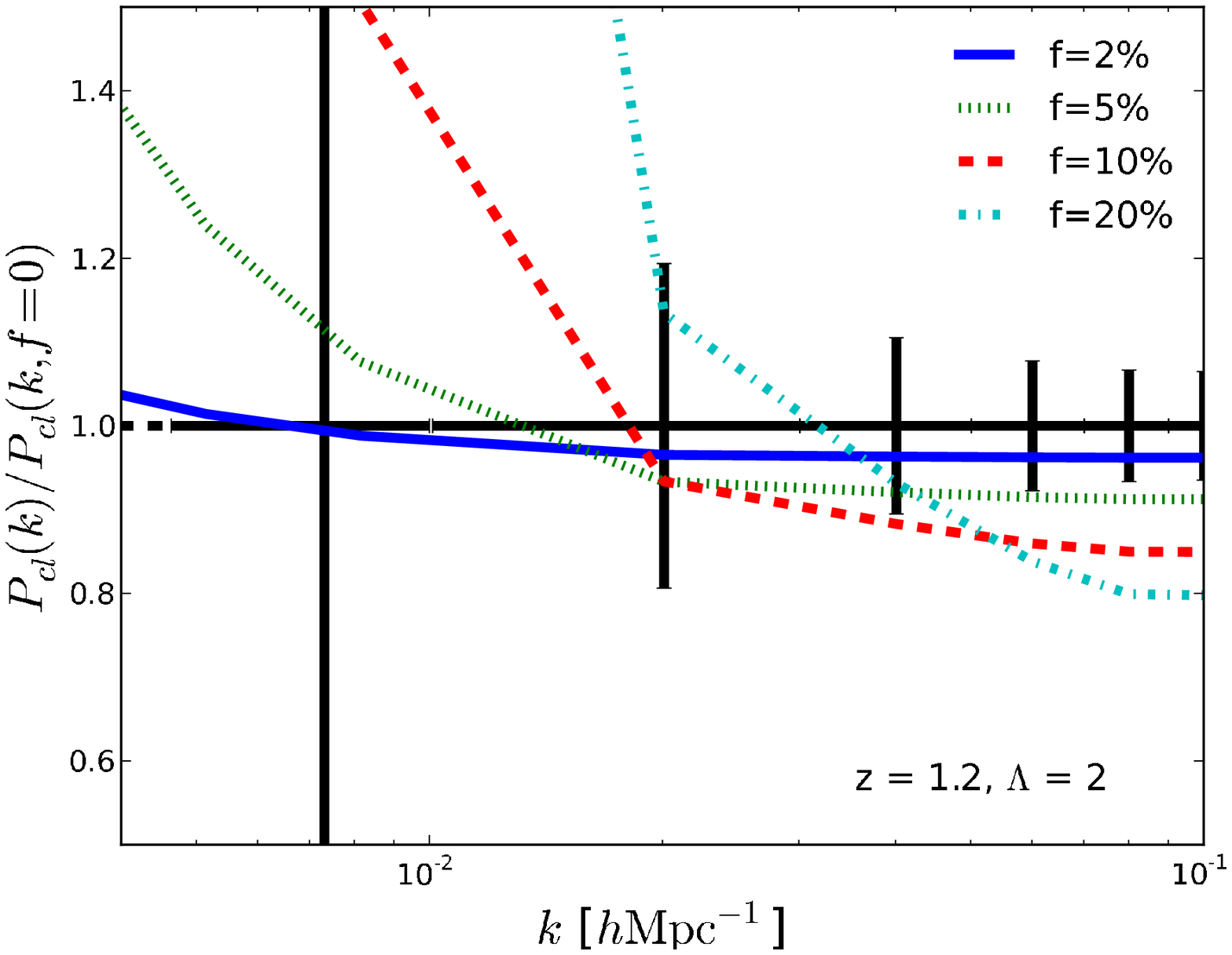}
\includegraphics[width=0.5\textwidth]{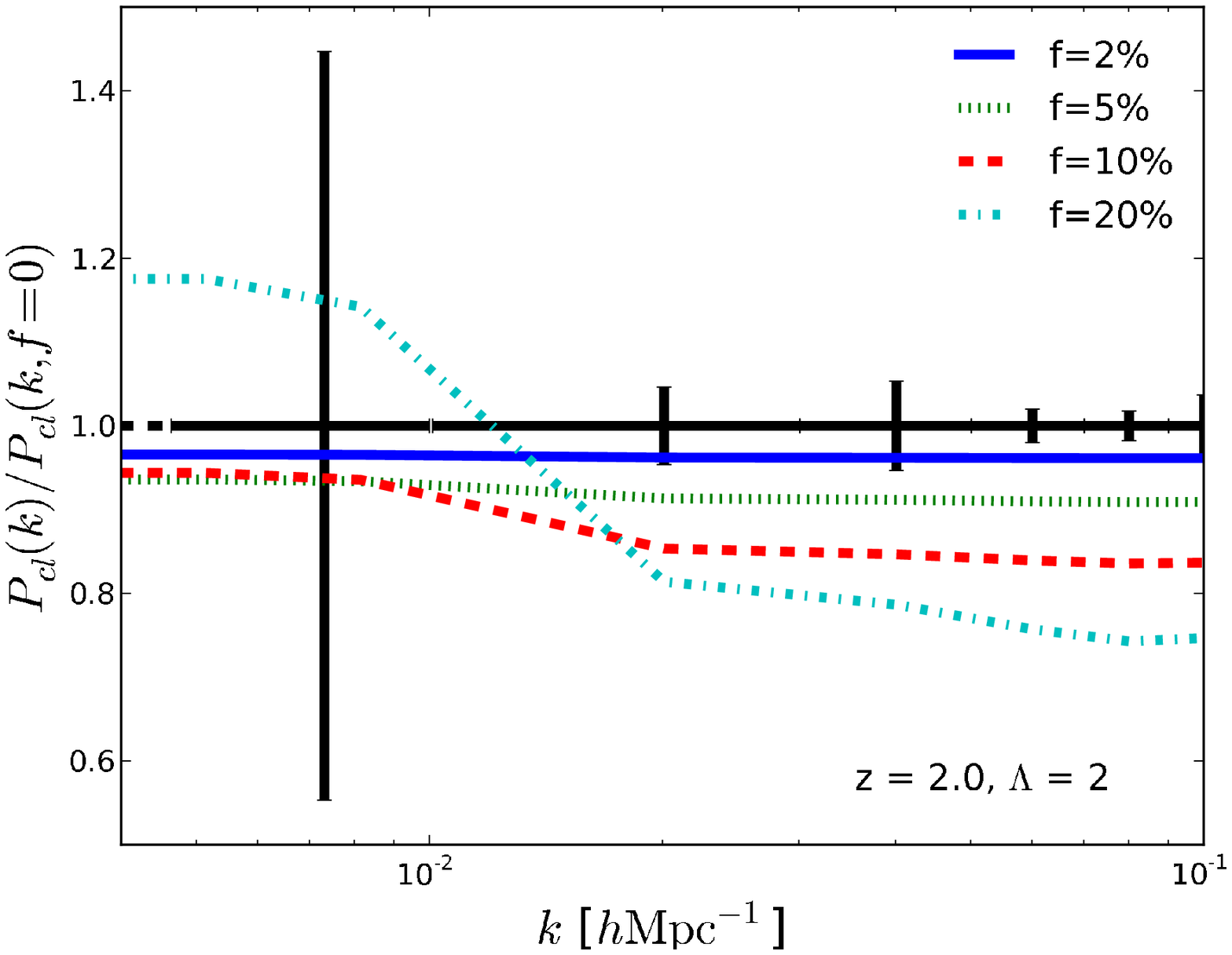}
\caption{\label{F:pkbias1} The predicted ratio between the measured 3D spherically-averaged galaxy power spectrum for $\Lambda=2$ (interlopers with longer wavelengths) assuming various levels of interloper contamination and the 3D power spectrum with no contaminations, along with 1$\sigma$ errors bars predicted for the PFS survey, where $\Lambda = \lambda_{\rm Int}/\lambda_{\rm SEL}$.  We plot the cases $f=2$\% (solid), $f=5$\% (dotted), $f=10$\% (dashed), and $f=20$\% (dot-dashed).  This plot shows that interlopers could significantly bias (up or down) power spectrum measurements for PFS.}
\end{center}
\end{figure}

\begin{figure}
\begin{center}
\includegraphics[width=0.5\textwidth]{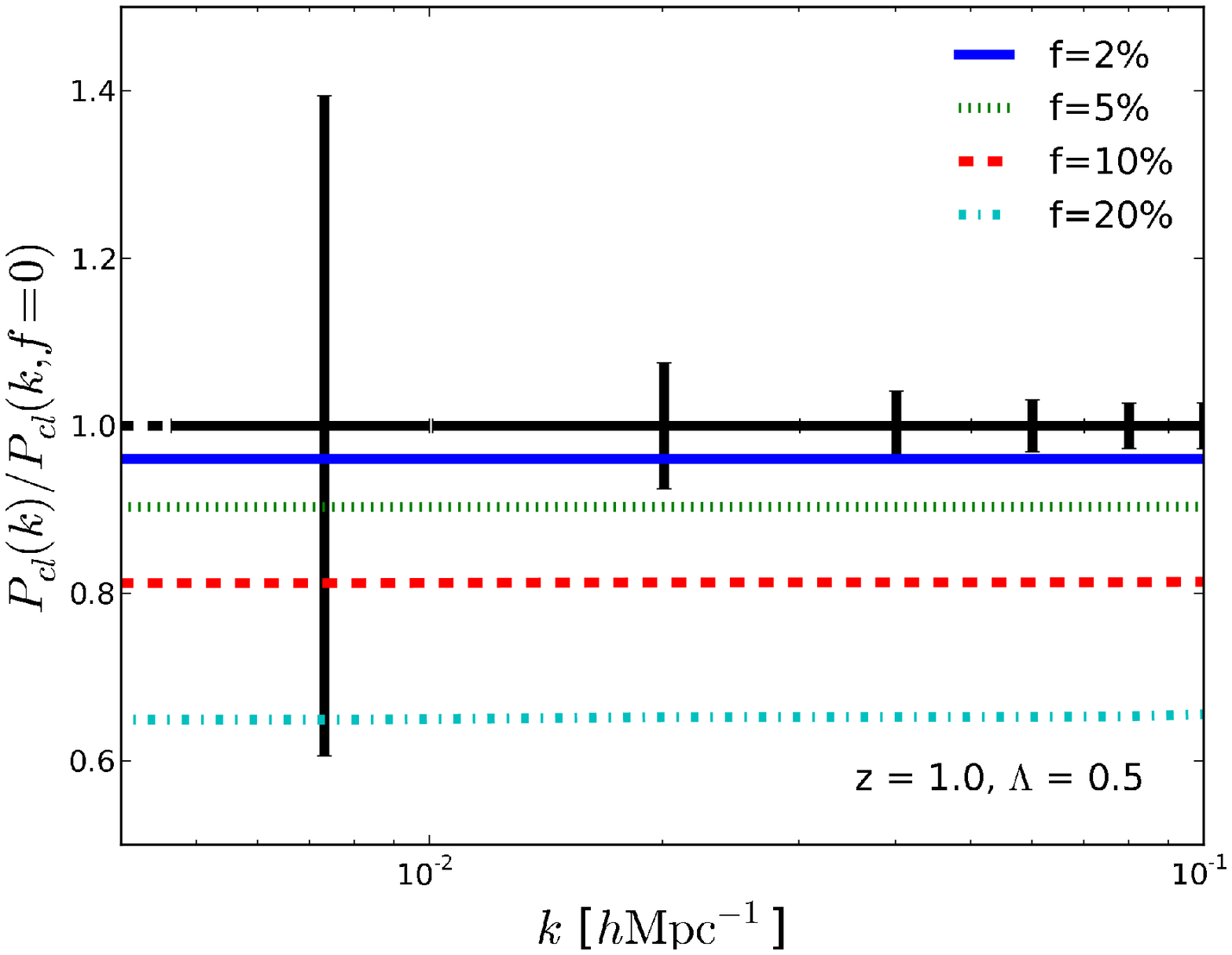}
\includegraphics[width=0.5\textwidth]{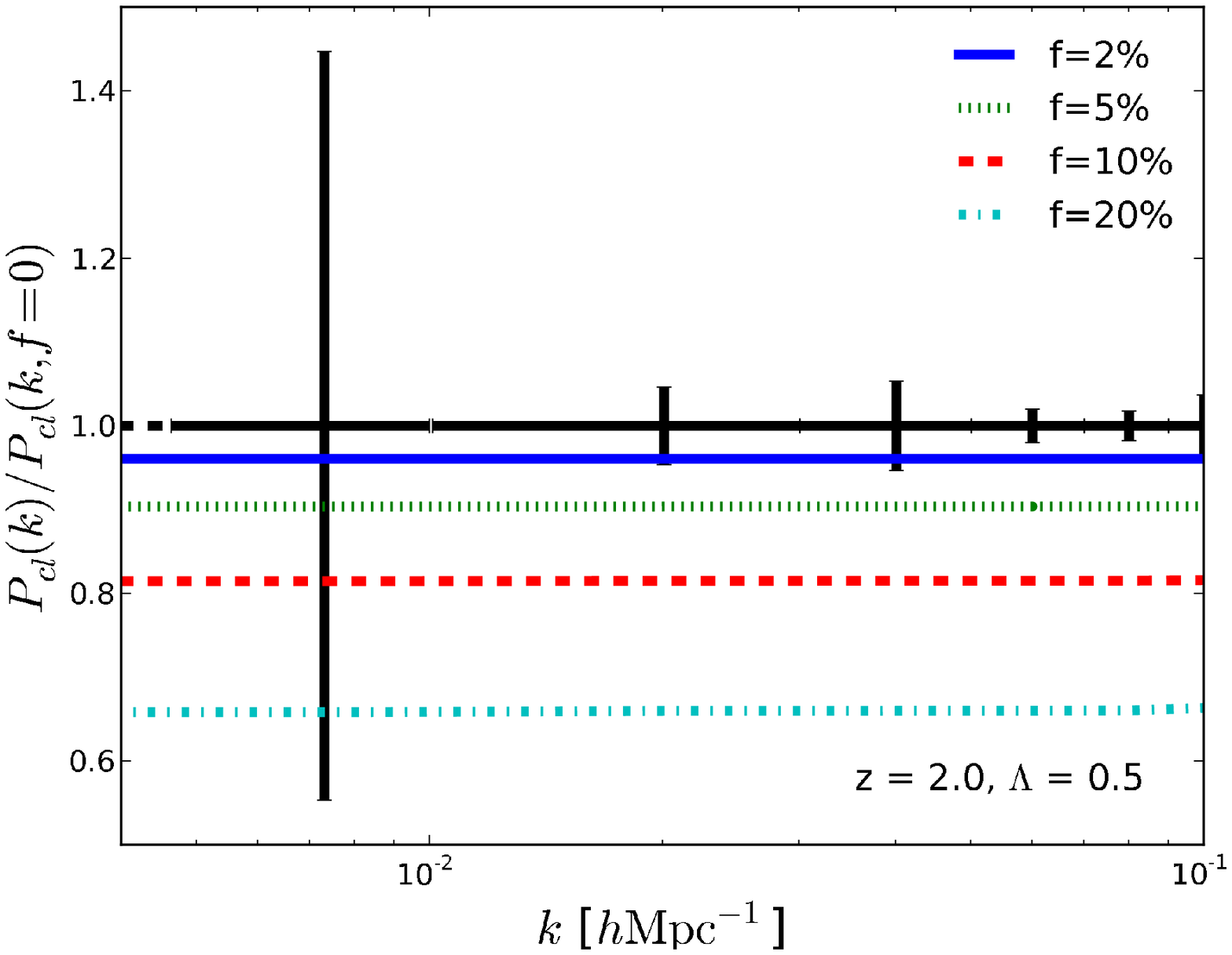}
\caption{\label{F:pkbias2} The predicted ratio between the measured 3D spherically-averaged galaxy power spectrum for $\Lambda=0.5$ (interlopers with shorter wavelengths) assuming various levels of interloper contamination and the 3D power spectrum with no contaminations.  The format is similar to Fig.~\ref{F:pkbias1}.  In this case, the volume distortion to the interloper power spectrum is small, causing the effect to the total power spectrum to be mainly a dilution of the true power spectrum.}
\end{center}
\end{figure}

In Figs.~\ref{F:pkbias1} and \ref{F:pkbias2}, we plot the biased, spherically averaged, linear clustering power spectrum due to interlopers for various redshifts and values of $\Lambda$ along with errors predicted for the PFS survey assuming no contamination.  For the clustering bias, we assume $b_g(z)=0.9+0.4z$, a fit \citep{2014PASJ...66R...1T} to semi-analytic models \citep{2010MNRAS.405.1006O} at the PFS flux limit that compares well with data.  According to this model, lower flux limits can increase $b_g$ by $\sim0.5$, and we find that the change in the fractional interloper distortion to the power spectrum due to increasing $b_g$ by one is much smaller than the PFS errors for small interloper fractions ($f\leq2$\%).  The same should be true for distortions to cosmological parameters.

In Table \ref{T:bnrbias}, we show the bias-to-noise ratio (BNR), which is the ratio of the interloper bias of $P(k)$ to its statistical error, averaged over the interval $k=10^{-4}-0.1h$Mpc$^{-1}$.  We find that an interloper contamination $f>0.25\%$ would bias the overall power spectrum signal relative to the $P(k)$ errors ($BNR>0.46$), such that the total measurement errors $\sigma_{\rm tot}=\sigma_P\sqrt{1+BNR^2}$ for PFS would increase by more than 10\%.  For WFIRST, this occurs if $f>0.2\%$.  For $f>0.5\%$ in either survey, we have $BNR>1$, which would be catastrophic for power spectrum measurements.  Thus, throughout our analysis $f<0.2\%$ will be our target interloper fraction; but note that if we have a significant interloper fraction that can be measured within $\pm0.2$\%, then we can accurately model the power spectrum of the SELGs well enough to account for it, particularly if the interloper contribution is small.  Note that the PFS and WFIRST designs have continued to evolve; we use the point designs in \citet{2014PASJ...66R...1T} and \citet{2013arXiv1305.5422S} but note that the analysis herein will have to be revisited for the final specifications.

Nonlinear clustering could in principle modify our predictions on small scales.  Several prescriptions exist for describing nonlinear clustering, including HALOFIT \citep{2003MNRAS.341.1311S}, convolution lagrangian perturbation theory \citep{2013MNRAS.429.1674C}, effective field theory \citep{2008JHEP...03..014C}, renormalization methods \citep{1998MNRAS.299.1097S}, and the halo model \citep{2006PhRvD..73f3519C,2002PhR...372....1C}.  In addition, nonlinear peculiar velocities can produce ``fingers of god'' (FoG) effects \citep{1972MNRAS.156P...1J} in the power spectrum.   More detailed work has been considered in the literature regarding nonlinear biasing \citep{2008PhRvD..78h3519M,2009ApJ...691..569J,2009JCAP...08..020M,2012PhRvD..86h3540B,2012PhRvD..86j3519C,2013MNRAS.433..209N,2014PhRvD..90l3522S,2015PhRvD..91b3508V} and nonlinear RSD \citep{2004PhRvD..70h3007S,2008PhRvD..78h3519M,2010PhRvD..82f3522T,2011MNRAS.417.1913R,2012JCAP...02..010O,2012JCAP...11..014O,2014MNRAS.439.3630W,2015arXiv150605814O}, yet here we will only consider a basic nonlinear clustering model.  Specifically, as an example we consider the nonlinear power spectrum using the HALOFIT prescription computed from CAMB \citep{2000ApJ...538..473L} along with a FoG damping term given in Eq.~10 of \citet{2011MNRAS.415.2876B} with $\sigma_v=2$ Mpc/$h$.  We find that the change in the fractional interloper distortion to the power spectrum due to nonlinear clustering is much smaller than the PFS errors for small interloper fractions ($f\leq2$\%).

For the specific purpose of constraining the effects of interlopers on BAO, we computed the BAO shifts for each of the cases in Table 1 (\emph{i.e.} each combination of $z_{SEL}$, $\Lambda$, and $f$). In each case, we computed the linear power spectrum, with interlopers, at each value of $\mu$ from 0 to 1 in steps of 0.1, and did a template fit as in \citet{2008ApJ...686...13S}, Eq.~1. The fits used a quadratic polynomial for $B(k)$, a 7th order polynomial for $A(k)$, and used the range of wavenumbers 0.02--0.3 $h$/Mpc. Fits were performed with shot noise levels appropriate for both $nP=0.2$ and $2$ (measured at $k=0.2h$/Mpc); see \citet{2015arXiv151003554B} Eq.~B3 for the explicit equation. The shift in the BAO scale is parameterized by $\alpha$, which re-scales the positions of the BAO features; $\alpha=1$ in the fiducial cosmology with no biases, but $\alpha>1$ ($\alpha<1$) indicates that the as-measured BAO ruler is shorter (longer) than the fiducial case. For the four combinations of $z_{SEL}$ and $\Lambda$, the largest shifts $|\alpha_{\rm with~interlopers} - \alpha_{\rm no~interlopers}|$ computed are at most 0.25\%\ for $f=0.02$, 0.06\%\ for $f=0.01$, 0.016\%\ for $f=0.005$, and 0.004\%\ for $f=0.0025$. Thus the BAO appears to be more robust against interlopers than the broadband power spectrum. The reason for this is that while interlopers dilute the power spectrum by a factor of $1-f$ (see Eq. 9), this dilution has no effect on the power spectrum shape; only the $f^2$ term coming from the clustering of the contaminants moves the BAO feature. It is thus both expected and numerically confirmed that the error in BAO measurements scales as $\propto f^2$. We also see that the BAO peak shift is negligible, even at the tenth-of-a-percent level of interest to future surveys, for $f<0.01$.

However, although a pure dilution (\emph{i.e.} adding unclustered fake sources into the sample) would also not effect the measured RSD parameter $\beta$, it would reduce the effective clustering bias of the sample.  This would in turn reduce the measured growth rate $f_g=\beta\times b_g$, which plays a large role in constraining modified gravity.  This is an example of the importance of a precise knowledge of the interloper fraction.

\begin{table}
\begin{center}
\caption{\label{T:bnrbias} The interloper bias-to-noise ratio (BNR) for P(k), marginalized over $k=10^{-4}-0.1h$Mpc$^{-1}$.  The first 4 rows assume PFS errors, while the last row assumes WFIRST errors.} 
\vspace{0.5cm}
\begin{tabular}{c|cccc}
\hline
$f$&$\Lambda=2$&$\Lambda=2$&$\Lambda=0.5$&$\Lambda=0.5$\\
&$z=1.2$&$z=2.0$&$z=1.0$&$z=2.0$\\
\hline
0.25\%&0.38&0.39&0.32&0.39\\
0.5\%&0.78&0.81&0.65&0.82\\
1\%&1.6&1.6&1.3&1.7\\
2\%&3.1&3.3&2.6&3.4\\
\hline
0.2\%&0.42&0.44&0.39&0.45\\
%
\hline
\end{tabular}\end{center}
\end{table}

\subsection{Cosmological Parameters}
In order to calculate the bias on cosmological parameters due to interloper contamination, let us call the measured power spectrum $\hat{d}_i(f)=P_t(f|k_i)$.  In this derivation, we will calculate a separate bias for each redshift bin, since the measurements in each redshift bin should be uncorrelated.  From this assertion, we can construct a chi-squared, assuming Gaussian errors, of the form
\begin{eqnarray}
\chi^2(\vecp)=[\estd(f)-\vecd(\vecp)]^T\matc^{-1}[\estd(f)-\vecd(\vecp)]\, ,
\end{eqnarray}
where $\vecp$ is a vector denoting the cosmological parameter values, $d_i(\vecp)$ is the predicted power spectrum with parameters $\vecp$, and $C_{ij}$ is the covariance matrix for power spectrum measurements at wavenumbers $k_i$ and $k_j$.  The model $\vecd(\vecp)$ can be expanded to first order from the fiducial cosmological parameters as
\begin{eqnarray}
d_i(\vecp)=d_{i,o}+J_{i\alpha}\Delta p_\alpha\, ,
\end{eqnarray}
where $J_{i\alpha}=\partial d_i/\partial p_\alpha$, $d_{i,o}=P_X(k_i)$, and $d_{i,o}$ and $J_{i\alpha}$ are evaluated at the fiducial parameter values.  However, since $\vecd(\vecp)\neq\estd$ for $f\neq0$, interloper contamination changes the location of the $\chi^2$ minimum, producing a bias $\Delta\vecp$ to any parameter estimations.  Note that this formalism is strictly true for small distortions in the power spectrum.  For large interloper fractions, our first-order expansion is insufficient and a Markov Chain Monte Carlo (MCMC) analysis is necessary to find the best-fit parameters upon interloper contamination.

By writing the change in the power spectrum, evaluated at the fiducial parameters, as $\Delta\vecd=\estd(f)-\vecd_o$, the resulting expression for $\chi^2$ is given by
\begin{eqnarray}
\chi^2(\vecp)=[\Delta\vecd-\mathbf{J}\Delta\vecp]^T\matc^{-1}[\Delta\vecd-\mathbf{J}\Delta\vecp]\, .
\end{eqnarray}
Optimizing this expression to find the minimum value, we find
\begin{eqnarray}
\mathbf{J}^T\matc^{-1}\mathbf{J}\Delta\vecp=\mathbf{J}^T\matc^{-1}\Delta\vecd\, .
\end{eqnarray}
Recognizing the expression for the Fisher matrix $\matf=\mathbf{J}^T\matc^{-1}\mathbf{J}$ and defining $\Delta\mathbf{D}=\mathbf{J}^T\matc^{-1}\Delta\vecd$, we solve for the parameter bias
\begin{eqnarray}
\Delta\vecp=\matf^{-1}\Delta\mathbf{D}\, .
\end{eqnarray}

While the matrix multiplications are strictly sums over wavevector bins, we can approximate them as integrals \citep{1997PhRvL..79.3806T,2003ApJ...598..720S}.  We use the same formalism as in \citet{2014PASJ...66R...1T}, where $\matf$ is given by their Eq.~(5) with no sum over redshift, and RSD effects are included in the power spectrum.  However, unlike in \citet{2014PASJ...66R...1T}, we use the full 3D power spectrum, not the observed $P(k_\parallel,k_\perp)$ model that is valid only for BAO fitting \citep{2003ApJ...598..720S}.  As is generally the case in Fisher analyses, we neglect correlations between bandpowers of the power spectrum due to nonlinear clustering or survey geometry, which slightly underestimates the errors. $\Delta\mathbf{D}$, similarly, is given by
\begin{eqnarray}
\Delta D_\alpha(z_i)
\!\!\!\!&=&\!\!\!\!\int_{-1}^1d\mu\,\int_{k_{\rm min}}^{k_{\rm max}}\frac{2\pi k^2dk}{2(2\pi)^3}V_{\rm eff}(k,\mu,z_i)\nonumber\\
&&\!\!\!\!\times\frac{\Delta P_t(f|k,\mu,z_i)}{P_{g,s}(k,\mu,z_i)}\frac{\partial\ln P_{g,s}(k,\mu,z_i)}{\partial p_\alpha}\nonumber\\
&&\!\!\!\!\times\exp\left[-k^2\Sigma_\perp^2-k^2\mu^2(\Sigma_\parallel^2-\Sigma_\perp^2)\right]\, ,
\nonumber \\ &&
\end{eqnarray}
using the notation of \citet{2014PASJ...66R...1T} where $P_{g,s}(k,\mu,z)=P_{\rm SELG}(k,\mu,z)$, $V_{\rm eff}$ is the effective volume, and the exponential factor accounts for the nonlinear BAO smearing.  The expressions and BAO smearing values for $V_{\rm eff}$, $\Sigma_\parallel$, and $\Sigma_\perp$ are given in Eqs.~6-8 of \citet{2014PASJ...66R...1T}.  We integrate over the wavenumber range $(k_{\rm min},k_{\rm max})=(10^{-4},0.5)\,h/$Mpc.  Although $k_{\rm max}=0.5h$/Mpc may seem a bit high, the exponential BAO smearing factor suppresses information naturally from smaller scales.  Additionally, in order to set conservative systematic requirements from interloper contamination, an ``optimistic'' estimate of statistical errors (\emph{i.e.}~large $k_{\rm max}$) should be chosen.  $\Delta P_t(f|k,\mu,z)$ corresponds to $\Delta\vecd$ and is just $\Delta P_t(f|k,\mu,z)=P_t(f|k,\mu,z)-P_{\rm SELG}(k,\mu,z)$.  We can also split $\Delta\mathbf{D}$ into two pieces according to
\begin{equation}
\Delta\mathbf{D} = [(1-f)^2-1]\Delta\mathbf{D}^0+f^2\Delta\mathbf{D}^{\rm Int}\, ,
\end{equation}
where
\begin{eqnarray}
\!\!\!\!\!\!\!\!
\Delta D_\alpha^0(z_i)\!\!\!\!
&=&\!\!\!\!\int_{-1}^1d\mu\,\int_{k_{\rm min}}^{k_{\rm max}}\frac{2\pi k^2dk}{2(2\pi)^3}V_{\rm eff}(k,\mu,z_i)\nonumber\\
&&\!\!\!\!\times \frac{\partial\ln P_{g,s}(k,\mu,z_i)}{\partial p_\alpha}\nonumber\\
&&\!\!\!\!\times\exp\left[-k^2\Sigma_\perp^2-k^2\mu^2(\Sigma_\parallel^2-\Sigma_\perp^2)\right]
\end{eqnarray}
and
\begin{eqnarray}
\Delta D_\alpha^{\rm Int}(z_i)\!\!\!\!&=&\!\!\!\!\int_{-1}^1d\mu\,\int_{k_{\rm min}}^{k_{\rm max}}\frac{2\pi k^2dk}{2(2\pi)^3}V_{\rm eff}(k,\mu,z_i)\nonumber\\
&&\!\!\!\!\times\frac{\gamma_\perp^2\gamma_\parallel P_{\rm Int}^{cl}[q(k,\mu),\mu_q(\mu),z_{\rm Int}(z_i)]}{P_{g,s}(k,\mu,z_i)}\nonumber\\
&&\!\!\!\!\times\frac{\partial\ln P_{g,s}(k,\mu,z_i)}{\partial p_\alpha}\nonumber\\
&&\!\!\!\!\times\exp\left[-k^2\Sigma_\perp^2-k^2\mu^2(\Sigma_\parallel^2-\Sigma_\perp^2)\right]\, .
\nonumber \\ &&
\end{eqnarray}
Using these expressions and an interloper fraction estimate, we can predict the exhibited bias for any cosmological parameter estimate per redshift due to interloper contamination for any LSS survey.

\subsection{Example: Growth Rate Bias}
In this section we focus on the effect of interlopers on measurements of growth rate parameters. In the formalism presented above (Eq.~\ref{E:pkmu}), the growth rate parameter $f_g$ enters in the RSD parameter $\beta$. 
The growth rate parameter $f_g$ is a key ingredient in our understanding of the correct cosmological model, as it is the logarithmic derivative of the linear growth factor, $D(a) \propto \delta_m$, with respect to the scale factor a:
\begin{equation}
f_g = \frac{d {\rm ln} D}{d {\rm ln} a} \, .
\end{equation}
In most cosmological and gravity models, $f_g$ can be parameterized as~\citep{2005PhRvD..72d3529L}:
\begin{equation}
\label{eq:fgamma}
f_g = \Omega_m(a)^\gamma \, ,
\end{equation}
where $\gamma$ is a parameter that is different for different cosmological models: in the standard $\Lambda$CDM+GR model it is a constant, $\gamma \approx 0.55$, while it is $\gamma \approx 0.68$ for the self-accelerating DGP model [see e.g.~\citet{2005PhRvD..72d3529L}]. In some other cases, it is a function of the cosmological parameters or redshift.

If one assumes that general relativity is the correct model for describing gravity, then the parameterization of Eq.~\ref{eq:fgamma} can be used to test cosmological parameters describing, \emph{e.g.} the dark energy equation of state.  In general, measuring $f_g$ gives strong constraints on the model of gravity, and it is the most popular measurement for testing general relativity and constraining modified gravity models [see e.g.~\citet{2008Natur.451..541G, 2008APh....29..336L, 2009MNRAS.393..297P, 2009JCAP...10..004S, 2010MNRAS.404..239S, 2011MNRAS.415.2876B, 2012PhRvD..85l3546Z, 2012MNRAS.420.2102S, 2013MNRAS.436...89R, Reid14, 2013MNRAS.433.1202S, 2013MNRAS.429.1514S, 2013arXiv1309.5385H, 2014MNRAS.439.3504S, 2014MNRAS.443.1065B, 2015arXiv150103821R, 2015arXiv150103840Z}].

Interlopers can in principle affect all cosmological parameters, in different ways. This also means that degeneracy between them in a large Fisher matrix analysis could change due to interloper effects. A detailed study of it is beyond the scope of the present paper, and we leave it to future work.

\begin{figure}
\begin{center}
\includegraphics[width=0.5\textwidth]{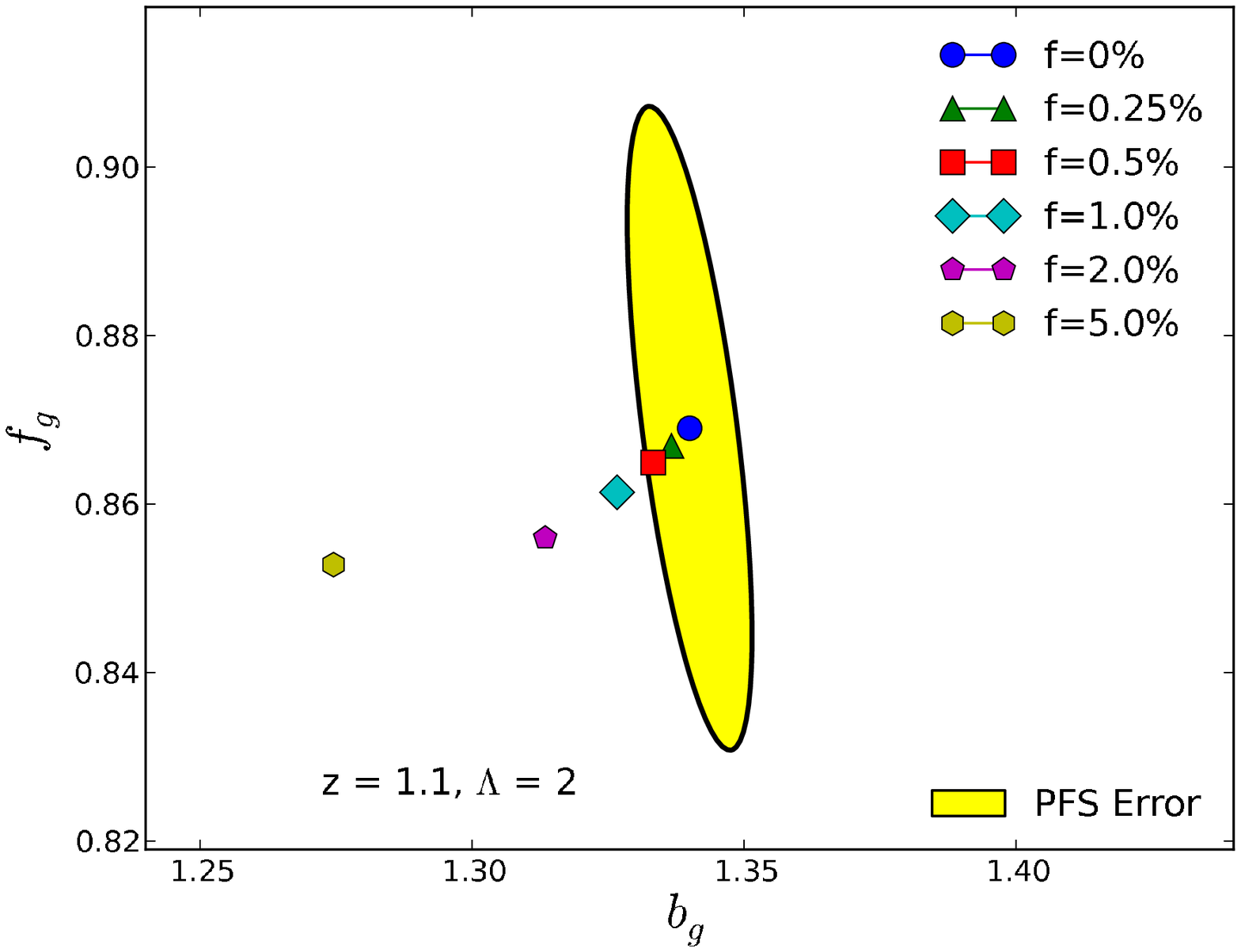}
\includegraphics[width=0.5\textwidth]{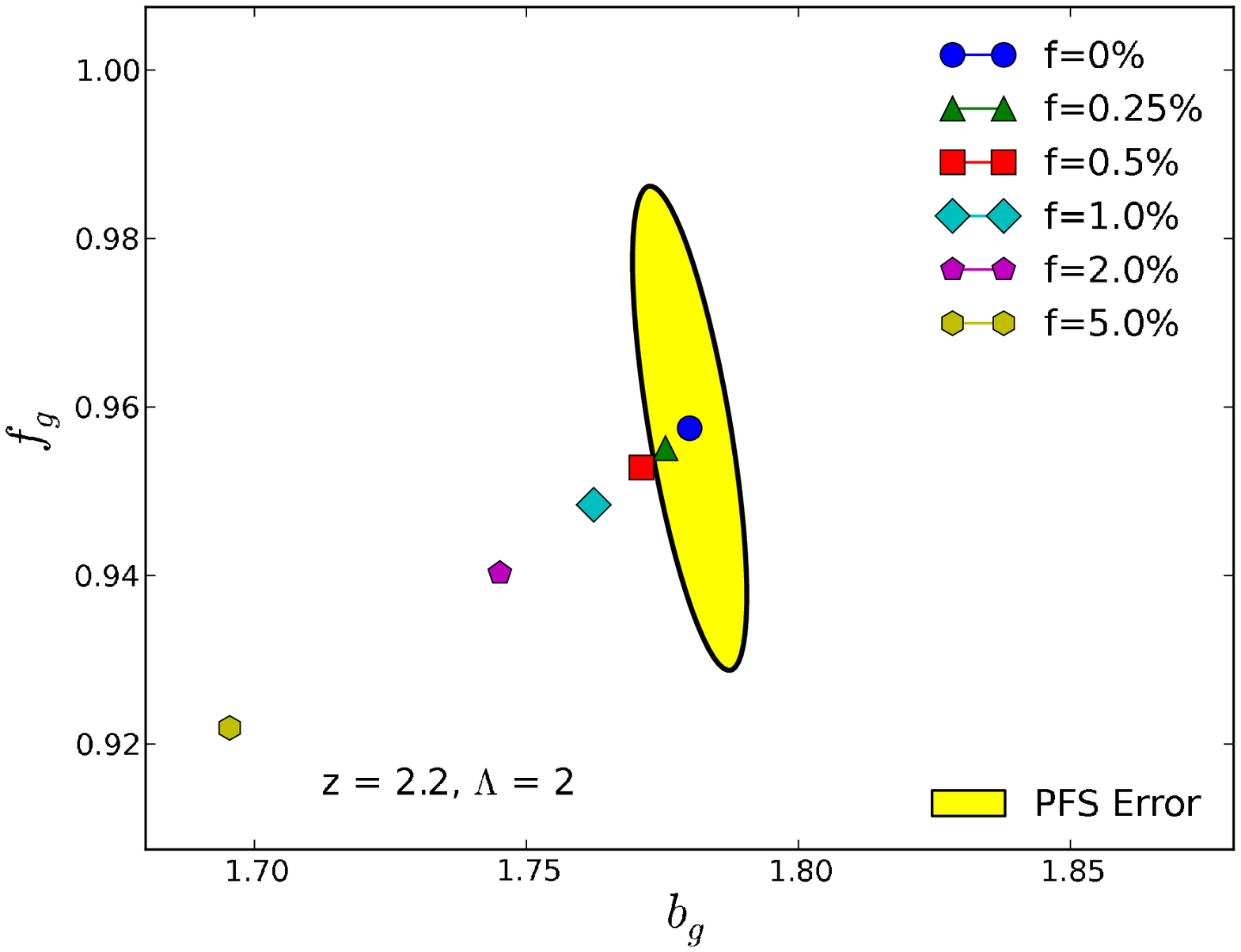}
\caption{\label{F:w0wa1} The shift in clustering bias and growth rate estimates due to interlopers for $\Lambda=2$ (interlopers with longer wavelengths), where $\Lambda = \lambda_{\rm Int}/\lambda_{\rm SEL}$.  We assume the fiducial parameters $(b_g,f_g)$ based on our fiducial model and mark the measured parameters for $f=0$ (circle), $f=0.25$\% (triangle), $f=0.5$\% (square), $f=1$\% (diamond), $f=2$\% (pentagon), and $f=5$\% (hexagon). We also show our predicted 1$\sigma$ PFS error ellipses for these parameters.  This plot shows that interlopers could significantly contaminate these parameter measurements for PFS.}
\end{center}
\end{figure}

\begin{figure}
\begin{center}
\includegraphics[width=0.5\textwidth]{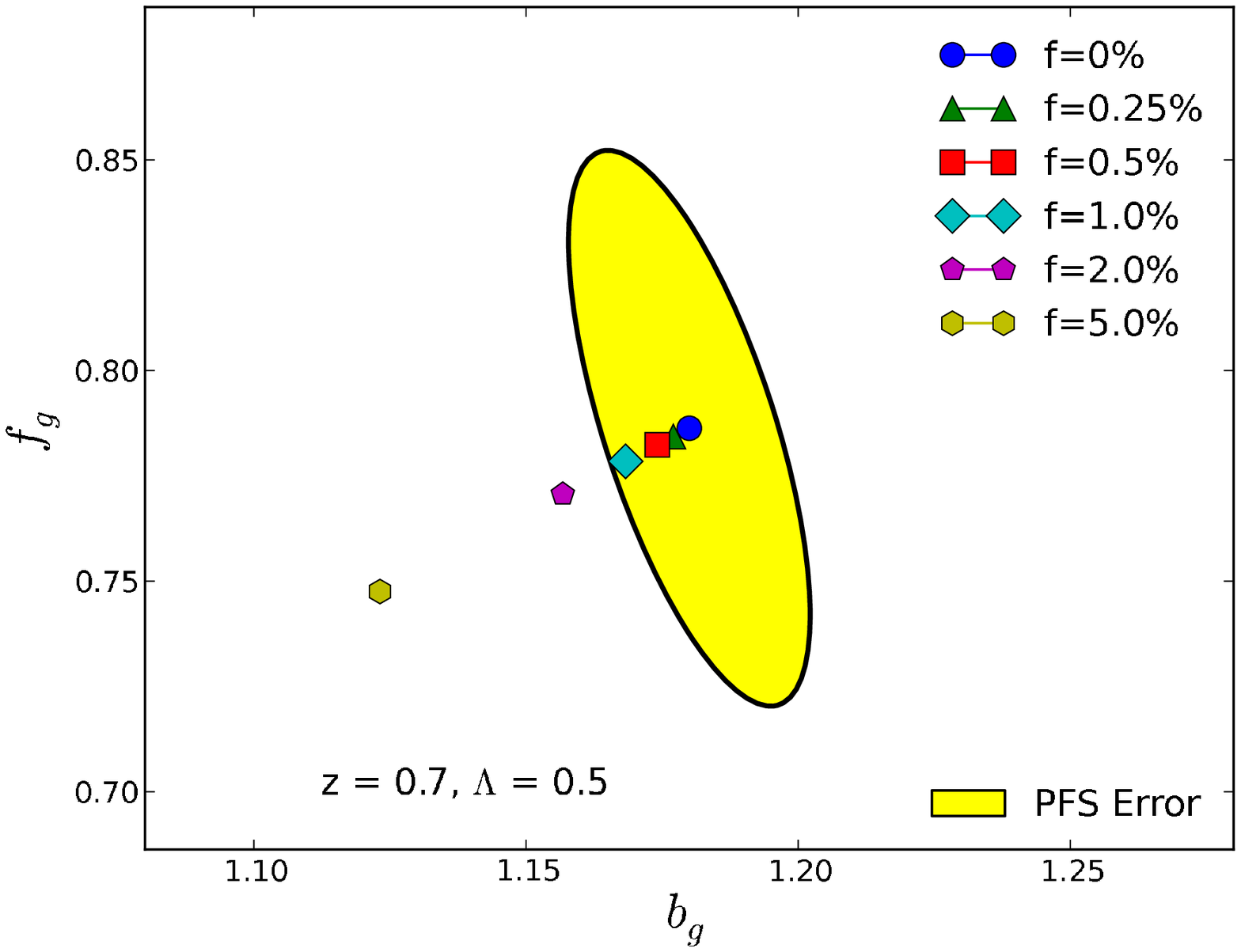}
\includegraphics[width=0.5\textwidth]{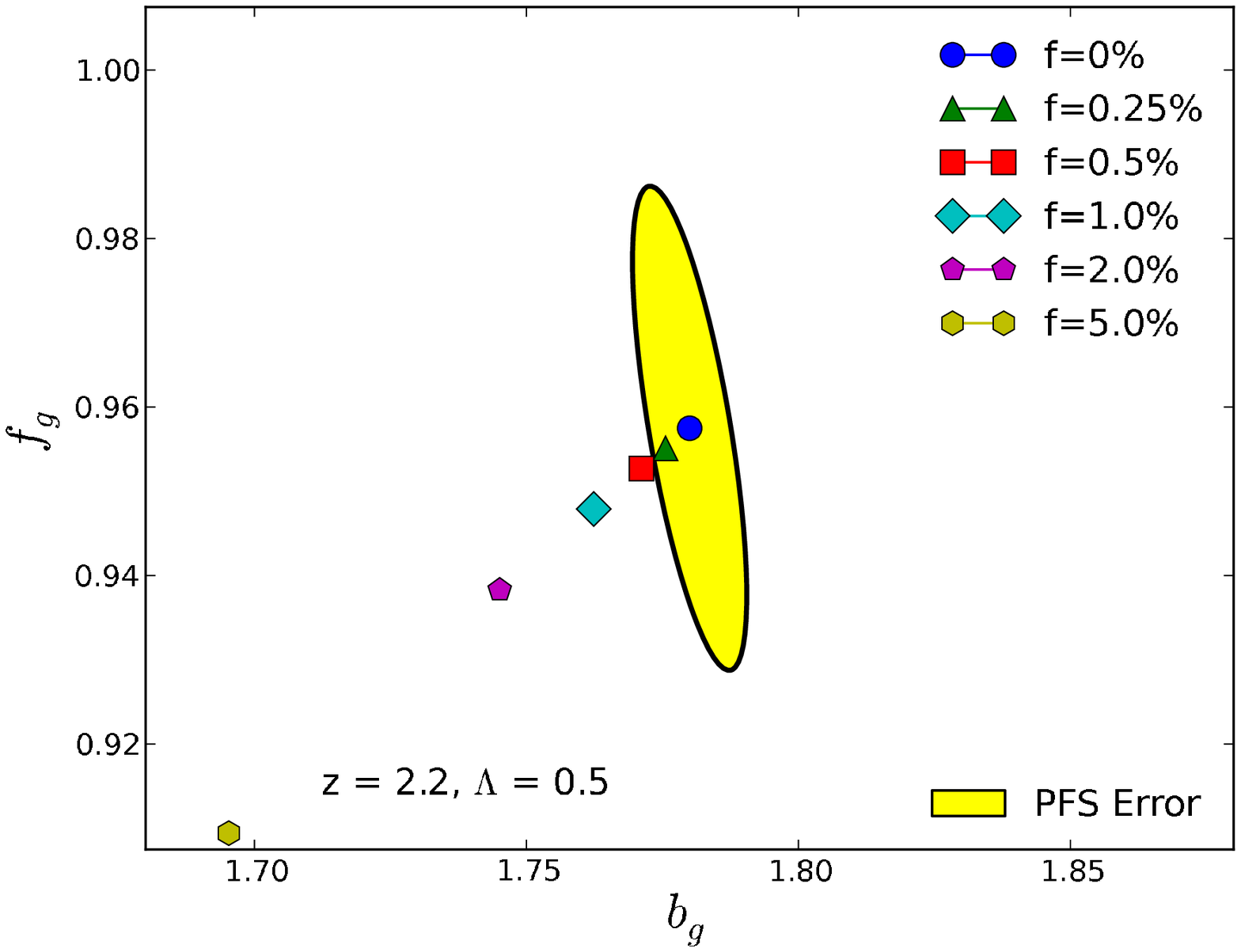}
\caption{\label{F:w0wa2} The shift in clustering bias and growth rate estimates due to interlopers for $\Lambda=0.5$ (interlopers with shorter wavelengths).  The format is similar to Fig.~\ref{F:w0wa1}.}
\end{center}
\end{figure}

We plot in Figs.~\ref{F:w0wa1} and \ref{F:w0wa2} the shift of the $(b_g,f_g)$ measurement due to the interloper fraction.  Note that we keep all other cosmological parameters constant, which underestimates the errors.  However, we do not expect this to affect our results because the $b_g$-$f_g$ degeneracy is much larger than degeneracies with other parameters.  Based on PFS errors, we can show that an interloper fraction $\gtrsim0.3$\% can bias the growth rate measurement such that the error on $f_g$ increases by more than 10\%. For WFIRST, the interloper fraction must be $\lesssim0.15$\% to not significantly bias growth rate measurements.  These target interloper fractions should be robust to nonlinearities and changes in the clustering bias, while correlations between band powers could increase the targets slightly.



\section{Methods to Remove Interlopers} \label{S:removal}

Since interlopers can greatly distort the power spectrum, future surveys will require methods to identify them. In principle this can be done at the level of individual galaxies, or through statistical means. There are various statistical methods for removing interlopers, including cross-correlation methods \citep{2008ApJ...684...88N,2013arXiv1303.4722M} and spectroscopic deep fields. Cross-correlation methods, where one correlates the survey map with another map with unambiguous redshifts, measure the product of bias$\times f$ without identifying the interloping galaxies. Spectroscopic deep fields would observe a representative sub-sample of the emission line galaxies to much higher signal-to-noise ratio -- and probably across a broader wavelength range in a multi-instrument campaign -- in hopes of detecting secondary lines for many more galaxies. We focus on methods that apply to individual galaxies here, and consider the statistical methods as a second line of defense that will be used to mitigate the residual interloper contamination in the catalog after galaxy-by-galaxy cleaning methods have been applied.


\subsection{Secondary Line identification} \label{S:2line}

One method to distinguish SELGs from interlopers is secondary line identification.  When a LSS survey measures an emission spectrum, it detects various atomic, ionic, and molecular emission lines.  Since these lines will all have the same redshift, the ratios of the line wavelengths with each other should be the same as the ratios in the rest frame.  Thus, we can identify a pair of emission lines by the ratio of their wavelengths {\it before we know the emitter's redshift}.  We can use this method to identify SELGs as well as interlopers.  This method is immune to contamination by other pairs of lines if no two sets of line pairs have the exact same wavelength ratio, which is true of the strong lines in the optical and near-infrared (NIR) spectral ranges we consider.\footnote{There are examples of line ratios that are almost the same; for example, He {\sc ii} 1640 : Ly$\alpha$ 1216 = H$\alpha$ 6563 : H$\beta$ 4861. However neither He {\sc ii} 1640 \AA\ nor Ly$\alpha$ is a significant interloper for far-red/NIR surveys at the depths being considered for PFS or WFIRST, as they would have to be at extraordinarily high redshifts.}

This method is limited for two reasons.  One is that a line $Z$ that works as a satisfactory secondary line for an emission line candidate at some redshifts will fall off the spectral range at other redshifts.  Another reason is that we cannot always detect every emission line.  If there was a line $\lambda_Z$ that always appeared with the survey emission line, then it would be necessary for any spectra with a SELG candidate to have a corresponding line $Z$ with the right wavelength.  However, not all emission lines will have a sufficient signal-to-noise ratio (SNR) to register as an ELG, especially when one of the lines intersects a sky line.  Thus, using this method will inevitably lead to some SELGs being rejected because its corresponding line $\lambda_Z$ did not appear in the spectrum.  It will also lead to interloping ELG candidates that are not SELGs being accepted because its corresponding line $\lambda_Z$ did not appear in order to rule it out as a SELG.  For each survey, we must assess which lines can consistently serve as secondary lines for SELGs and their interlopers, as well as determine how often the secondary line test fails as a function of observed wavelength, or equivalently of $z_{\rm SELG}$.

For secondary lines, we will take the conservative approach by requiring the secondary line to have a SNR high enough to prevent a statistical fluctuation from masquerading as a secondary line.  To determine the necessary SNR, we compute the probabilities $P_1$ and $P_2$, where $P_1=P({\rm not\,accepted|real})$ is the probability that a real line is rejected because it does not exceed our chosen SNR cutoff and $P_2=P({\rm accepted|not\,real})$ is the probability that a statistical fluctuation in the spectrum is accepted as a real line because the fluctuation was higher than the SNR cutoff.  Assuming Gaussian fluctuations from an expected flux $F_e$ and a flux cutoff $F_c$, we find the two probabilities are given by
\begin{eqnarray} \label{E:secsnr}
P_1({\rm not\,accepted|real}) \!\!\!\!&=& \!\!\!\! \frac{1}{2}\left[1-{\rm erf}\left(\frac{Q_e-Q_c}{\sqrt{2}}\right)\right],\nonumber\\
P_2({\rm accepted|not\,real}) \!\!\!\!&=& \!\!\!\! \frac{1}{2}\left[1-{\rm erf}\left(\frac{Q_c}{\sqrt{2}}\right)\right]\, ,
\end{eqnarray}
where $F_n$ is the flux noise of the instrument, $Q_e=F_e/F_n$, and $Q_c=F_c/F_n$.  As expected, increasing the cutoff $Q_c$ increases $P_1$, the rate of rejected true lines, and decreases $P_2$, the rate of accepted false lines.  In general, we require secondary lines to have a SNR greater than $Q_c=4$, making the number of statistical fluctuations accepted as true lines negligible at the expense of eliminating a fair amount of true lines.

We relax the required SNR for the secondary line \oiii\ 4959 when identifying \oiii\ 5007 to $Q_c=1$ ($P_2=16\%$; see Eq.~\ref{E:secsnr}) since we know the line ratio $F_e(5007{\rm \AA}):F_e(4959{\rm \AA})=3$ from atomic physics \citep{2000MNRAS.312..813S}. Setting the \oiii\ 5007\AA\, line as the primary line, we can set the minimum value for $Q_e$ for the secondary line \oiii\ 4959\AA\, equal to
\begin{eqnarray}
Q_{e,{\rm min}} = \frac{SNR_{\rm min}(5007{\rm \AA})F_n(5007{\rm \AA})}{3F_n(4959{\rm \AA})}\, .
\end{eqnarray}
It should be noted that when the 5007\AA\, line is the interloper, secondary line identification will cause the number counts of the \oiii\ 5007\AA\, interloper to be multiplied by a factor of $P_1$ and the number counts of \oii\ emitters to be multiplied by $(1-P_2)$.  However, when the 5007\AA\, line is the SELG, as in WFIRST, secondary line identification will cause the number counts of interlopers to be multiplied by a factor of $P_2$ and the number counts of \oiii\ lines to be multiplied by $(1-P_1)$.  Thus, since $P_1$ and $P_2$ are anti-correlated, our goal for each case is to set $Q_c$ such that the first probability is minimized without decreasing the second probability so much that the shot noise from the SELGs gets too big.

\subsection{Photometry}

We can also use photometry to rule out candidates for SELGs.  As the spectrum of an object redshifts, its colors, or brightness differences between neighboring photometric bands, will traverse through color space.  Thus, an object's photometric colors can determine its \emph{photometric redshift}.  Photometric redshift samples have been constructed in numerous surveys, e.g. the Sloan Digital Sky Survey (SDSS) \citep{2011ApJ...728..126W}, the Canada-France-Hawaii Telescope Legacy Survey (CFHTLS) \citep{2012MNRAS.421.2355H}, COSMOS \citep{2009ApJ...690.1236I}. Techniques to determine photometric redshifts will also be used in weak lensing surveys such as Dark Energy Survey (DES) \citep{2010JPhCS.259a2080S}, Hyper Suprime-Cam (HSC)\footnote{http://sumire.ipmu.jp/en/2652} and LSST \citep{2012arXiv1211.0310L}, as well as \emph{Euclid} and WFIRST.

In this study, we do not attempt to construct a scheme for determining photometric redshifts.  Instead, we determine photometric cuts that separate SELGs from interlopers.  As an example, let us consider an \oiii\ emission line contaminating an H$\alpha$ survey.  The \oiii\ interlopers will come from a higher redshift than the H$\alpha$ galaxies, which imply that for a small enough redshift range, they may inhabit different locations in color space, or different color loci.  If this is the case, then a photometric cut that separates the two color loci can be determined, and this photometric cut can be applied to SELG candidates in the survey to identify interlopers.  The redshift ranges used for this method must be small to prevent the color loci from overlapping.  In our analysis we determine photometric cuts within two-dimensional slices of color space, although it is possible to reach better accuracy by determining cuts within the full multi-dimensional color space.


\section{Mock Surveys} \label{S:mock}

We assess the performance of the secondary line identification and photometry methods in reducing the interloper rate $f(\lambda_{\rm SEL}-\lambda_{\rm Int},z_{\rm SELG})$, the fraction of galaxies identified in a survey using emission line $\lambda_{\rm SEL}$ that are actually interlopers with emission line $\lambda_{\rm Int}$, given by
\begin{eqnarray}
&&f(\lambda_{\rm SEL}-\lambda_{\rm Int},z_{\rm SELG})\nonumber\\
&&=\frac{N_{\rm Int}(z_{\rm Int})}{N_{\rm SELG}(z_{\rm SELG})+N_{\rm Int}(z_{\rm Int})}\, ,
\end{eqnarray}
where $N_{\rm SELG}(z_{\rm SELG})$ is the number of galaxies at redshift $z_{\rm SELG}$ identified by emission line $\lambda_{\rm SEL}$, and $N_{\rm Int}(z_{\rm Int})$ is the number of galaxies at redshift $z_{\rm Int}$ satisfying Eq.~\ref{E:zint} such that its emission line $\lambda_{\rm Int}$ interlopes the SEL.  This definition of $f$ is equivalent to Eq.~\ref{E:fn}.  We seek interloper rates for three cases: (1) we include all interlopers with fluxes great enough to pass the survey's detection criterium, (2) we include all interlopers from case (1) that also fail to be identified by the survey's secondary line identification test for the interloping ELG, and (3) we include all interlopers from case (2) that fail to be identified using photometry.

We evaluate $f(\lambda_{\rm SEL}-\lambda_{\rm Int},z_{\rm SELG})$ by tabulating number counts from the COSMOS \citep{2007ApJS..172...99C,2009ApJ...690.1236I} Mock Catalog (CMC) \citep{2011A&A...532A..25J} for each potential interloper.  The COSMOS survey is a combination of various surveys which, after cutting out areas masked due to bright stars, together include 538,000 galaxies over 1.24 deg$^2$.  The CMC was constructed by converting the UV magnitudes of the galaxies to star formation rates to \oii\ luminosities using the Kennicutt relations \citep{1998ARA&A..36..189K}.  The SFR-UV calibration is based on the Salpeter stellar initial mass function (IMF) \citep{1955ApJ...121..161S}, which would be modified for more modern IMFs, \emph{e.g.} \citet{2003PASP..115..763C}, Also, the error on the \oii\-SFR calibration is 30\%, mainly due to the different values for different stellar types.  The \oii\ luminosities are then converted to other emission lines, including H$\alpha$, H$\beta$, and \oiii\ , using measured line ratios, many of which are uncertain and out-of-date.  Thus we do not expect the CMC to perfectly reflect reality, but it serves as a useful first look at expected interloper rates for future surveys.

We also recalibrate the CMC to account for updated luminosity functions (LFs).  Note that many of the emission lines were recently updated (Zoubian et al., in prep.). The H$\alpha$ lines in the CMC were calibrated to the LF in \citet{2010MNRAS.402.1330G}.  The other line luminosities were scaled from H$\alpha$ using measured line ratios \citep{2009ApJ...690.1236I}.  Thus \oiii\ luminosities are thus also calibrated to the H$\alpha$ LF, but this is particularly dangerous given the large observed variation in \oiii/H$\alpha$ ratios.  We recalibrate the H$\alpha$ and \oiii\ lines to be consistent with the recent LFs from \citet{2013ApJ...779...34C}.  Specifically, we use the Colbert LF for \oiii\ with $\alpha$ fixed.  Also note that the \oiii\ LF in \citet{2013ApJ...779...34C} applies to the \oiii\ 5007 flux only, not the total doublet flux.  The recalibration is performed for each line by first comparing the number densities $n(>L)$ based on the old and new LF then transforming each emission line galaxy's luminosity such that $n(>L)$ matches the new LF.  Specifically, we compare $n(>L_{\rm H\alpha})$ from \citet{2010MNRAS.402.1330G} to $n(>L_{\rm H\alpha})$ and $n(>L_{\rm OIII})$ from \citet{2013ApJ...779...34C} to recalibrate H$\alpha$ and \oiii\ 5007.  We performed these recalibrations in redshift bins of $\Delta z=0.1$.  We also recalibrate \oiii\ 4959 by setting the \emph{intrinsic} line ratio $L_{5007}/L_{4959} = 3$.

\begin{figure}[!t]
\begin{center}
\includegraphics[width=0.5\textwidth]{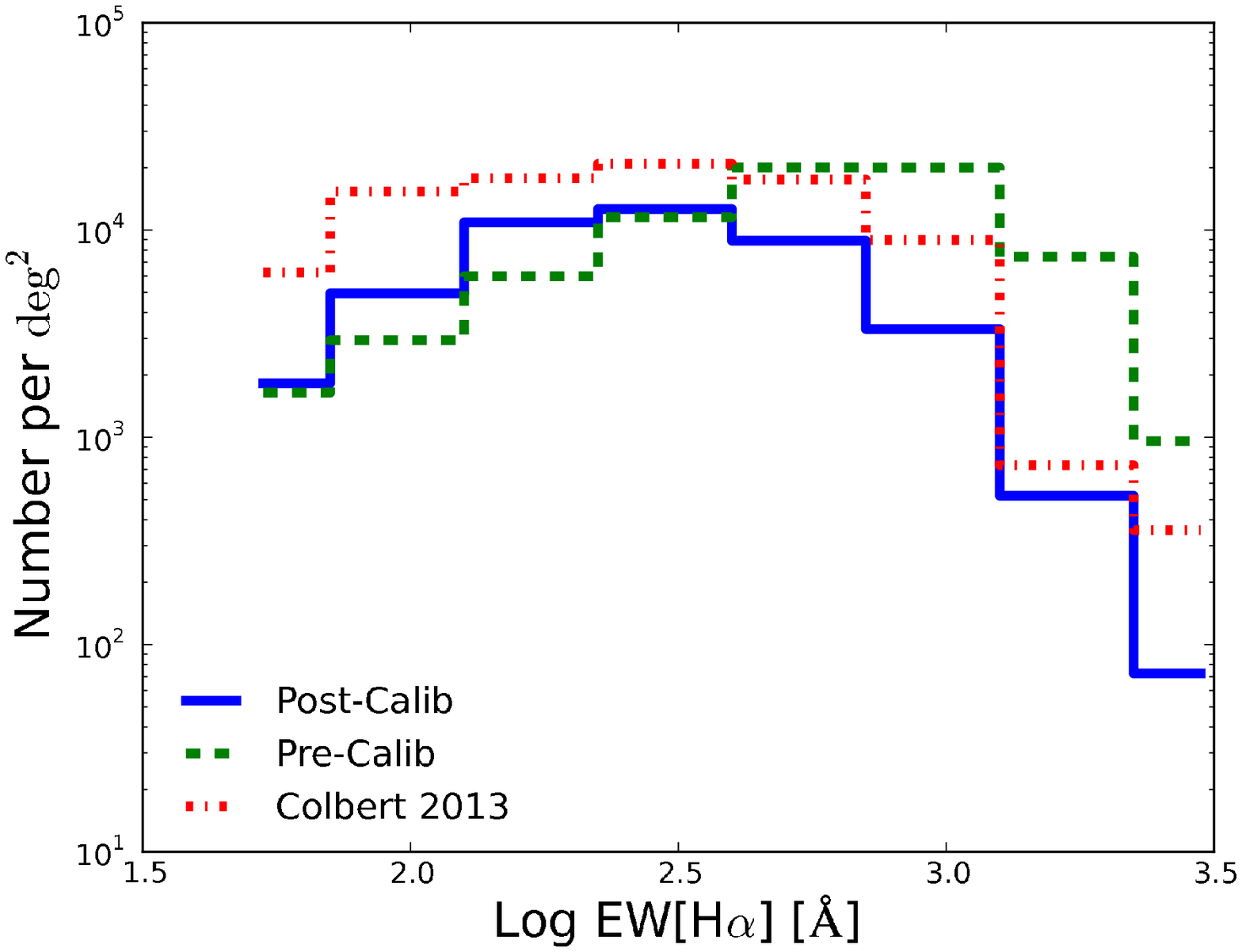}
\includegraphics[width=0.5\textwidth]{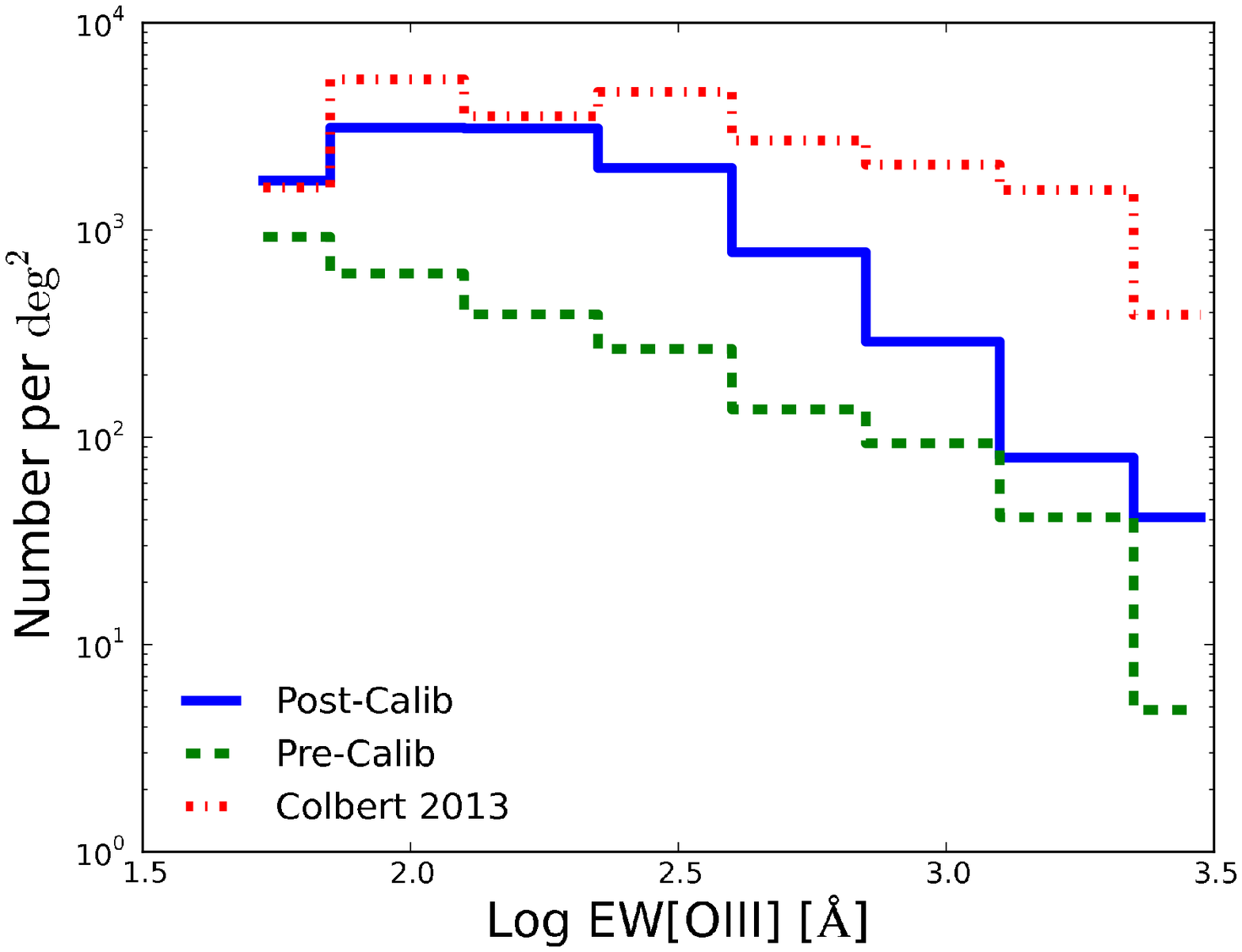}
\caption{\label{F:ew} The equivalent width (EW) distributions of H$\alpha$ (top) and \oiii\ (bottom) emitters in a mock WISP sample constructed from the CMC.  The dashed (solid) histograms are the distributions before (after) re-calibration to LFs in \citet{2013ApJ...779...34C}.  The dash-dotted histograms are from \citet{2013ApJ...779...34C}.  The EW distributions from the re-calibrated sample show better agreement with \citet{2013ApJ...779...34C} than the un-calibrated sample, yet there are differences, particularly in the \oiii\ sample at high EW.}  Note that plots in \citet{2013ApJ...779...34C} and our plot include $F_{\rm[N\,{\sc ii}]}=0.4F_{\rm H\alpha}$ in the H$\alpha$ flux, while \oiii\ in \citet{2013ApJ...779...34C} and our plot refers to the \oiii\ 5007 line only.
\end{center}
\end{figure}

In order to show that our re-calibrated emission lines indeed have emission properties consistent with \citet{2013ApJ...779...34C}, we plot equivalent width (EW) distributions for H$\alpha$ and \oiii\ emitters within a mock sample produced using the CMC corresponding to the HST WFC3 Infrared Spectroscopic Parallels (WISP) survey \citep{2010ApJ...723..104A}, which was used to determine the luminosity functions in \citet{2013ApJ...779...34C}.  Note that, similar to \citet{2013ApJ...779...34C}, we require the H$\alpha$ and \oiii\ lines to have a $SNR>5$, where the spectral noise is given in Fig.~5 of \citet{2010ApJ...723..104A} according to the WISP survey.  We plot the distributions both before and after re-calibration.  In Fig.~\ref{F:ew}, we see better consistency after re-calibration than before with the EW distributions in Figs.~5 and 6 of \citet{2013ApJ...779...34C}.  However, we do see significant differences for the \oiii\ sample at high redshifts.  Assuming our calibrations were performed correctly, this would imply that the continuum measurements in the mock catalog are not consistent with those from the WISP survey.  These results suggest that approximately 40\% of galaxies may be affected.

As another check, we also compare the intrinsic H$\alpha$/H$\beta$ line ratio before and after re-calibration, taking into account the galactic extinction.  Before re-calibration, we find the mean line ratio to be 3.70, while after re-calibration, the ratio decreases to 2.92, which is much closer to the expected value of 2.86 based on atomic physics predictions \citep{2003adu..book.....D}.

\begin{figure}[!t]
\begin{center}
\includegraphics[width=0.5\textwidth]{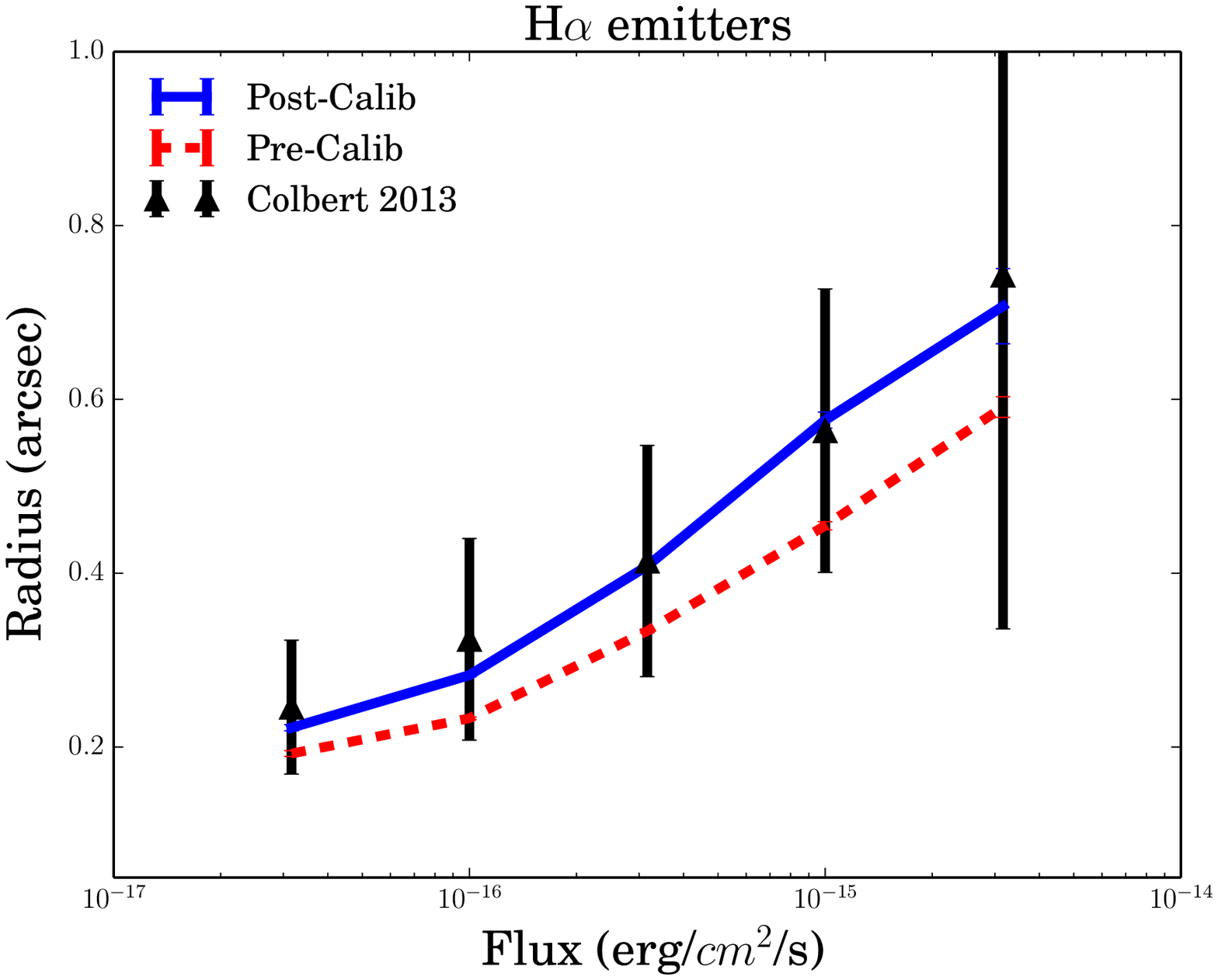}
\includegraphics[width=0.5\textwidth]{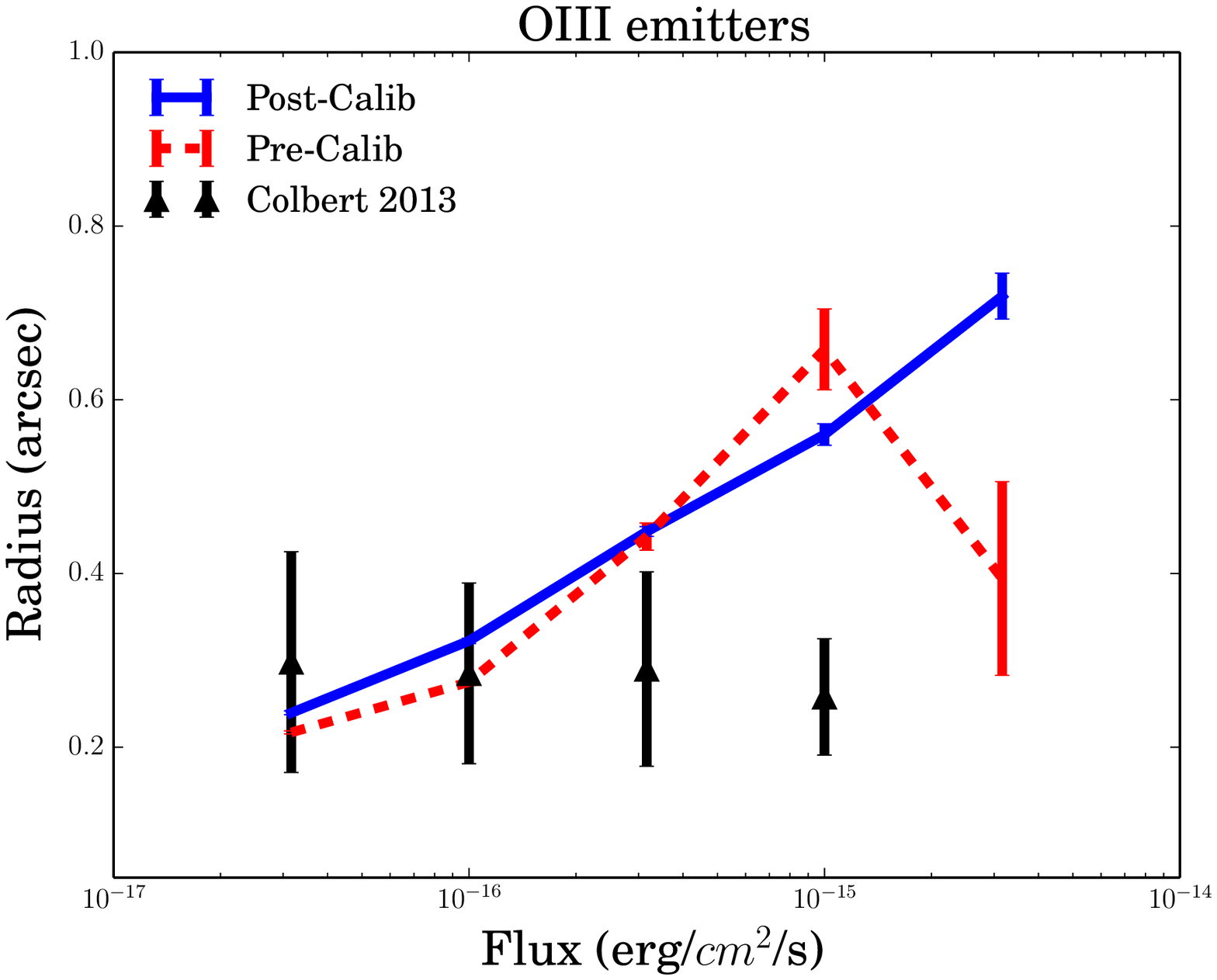}
\caption{\label{F:radflux} The radius-flux relation for H$\alpha$ (top) and OIII (bottom) emitters in a mock WISP sample constructed from the CMC.  The dashed (solid) points are the distributions before (after) re-calibration to LFs in \citet{2013ApJ...779...34C}.  The black triangles are the points from \citet{2013ApJ...779...34C}.  The radius-flux relation derived from the re-calibrated sample mostly show agreement with \citet{2013ApJ...779...34C}.  Note that plots in \citet{2013ApJ...779...34C} and our plot include $F_{\rm NII}=0.4F_{\rm H\alpha}$ in the H$\alpha$ flux, while \oiii\ in \citet{2013ApJ...779...34C} and our plot refers to the \oiii\ 5007 line only.}
\end{center}
\end{figure}

We also plot in Fig.~\ref{F:radflux} the average radius binned over H$\alpha$ and \oiii\ flux for all the objects in this mock WISP sample, both before and after re-calibration, in an attempt to reproduce Fig.~11 in \citet{2013ApJ...779...34C}.  Our error bars are much smaller than those in their Fig. 11 because we include objects over the whole COSMOS field, which is $\sim37\times$ larger than the WISP field.  Our results are mostly consistent with their estimates of the radius-flux relations for the two emitters except for the 4th flux bin for the \oiii\ emitters, in which our average radius is higher than their estimate by about 3$\sigma$.  However, our \oiii\ radius-flux relation after calibration is closer than the pre-calibration result to \citet{2013ApJ...779...34C}, and this discrepancy should not affect our results since \oiii\ emitters at such high fluxes should be easily observed by PFS and WFIRST, regardless of the effective radius.

Recently, the \emph{Cosmic Assembly Near-Infrared Deep Extragalactic Legacy Survey} (CANDELS) \citep{2013ApJS..206...10G,2013ApJS..207...24G} released 0.1 deg$^2$ of multi-wavelength data.  This data revealed that the CMC has an excess of galaxies with photometric measurements $i-H>1$ not present in the CANDELS data.  This discrepancy is mostly due to extrapolations of photometry for very faint objects, as well as possibly incorrect spectral and photometric templates.  We find that many galaxies in the CMC that comprise the PFS and WFIRST surveys are given $i$ and $H$ band measurements with $i-H>1$.  Specifically, 29\% of PFS galaxies, 92\% of WFIRST H$\alpha$ galaxies, and 89\% of \oiii\ galaxies in the CMC are in this region in color space, implying that our photometric tests and simulations for these surveys may be too optimistic (i.e. based on simulated galaxies that have stronger Balmer breaks than the real galaxies). Fixing this discrepancy between the mock catalog and the observed population of galaxies will be the subject of ongoing work, as the mock catalogs used in LSS survey forecasting have undergone continuous improvement over the years. For the purposes of this paper, we merely note the existence of this issue and comment later on its possible implications.

\subsection{PFS mock survey}

The PFS survey, which has a spectral range of $0.67{\rm \mu m}<\lambda<1.26{\rm \mu m}$, will determine the spectroscopic redshifts of \oii\ galaxies in the redshift range $0.8<z<2.38$.  The \oii\ doublet consists of two emission lines with wavelengths 3726\AA ~and 3729\AA.  The PFS spectral resolution is high enough to resolve both lines if they are visible, but sky lines and the variable line ratio of the \oii\ doublet could cause one of the peaks to not be visible.  Therefore, we must be able to tell if a single line could possibly be \oii\ emission rather than other emission lines, or interlopers.

In this analysis, we determine the rate at which interlopers will appear as \oii\ emitters in PFS, as well as the performance of other strategies to minimize these interloper rates.  The interlopers we consider in this study are Ly$\alpha$ 1216, H$\beta$ 4861, \oiii\ 4959/5007, H$\alpha$ 6563, \nii\ 6584, and \sii\ 6727/6731.  Note that PFS will have a spectral resolution high enough to resolve between H$\alpha$ and \nii, as well as the \oiii\ doublet.  The PFS survey will use HSC for target selection of \oii\ candidates.  PFS will only target emitters that pass the following photometric cuts,
\begin{eqnarray} \label{E:pfstargets}
22.8\leq g\leq 24.2\,{\rm AND} -0.1<g-r<0.3\nonumber\\
{\rm AND\,NOT}\,(g>23.6\,{\rm AND}\,r-i>0.3)\, .
\end{eqnarray}
We simulate these cuts for the \oii\ emitters and the interlopers in the CMC.  Fluctuations in the magnitudes due to instrumental noise could affect which galaxies pass the cuts.  We account for this by converting the CMC magnitudes to continuum fluxes $f_\nu$ for all the galaxies, upon which we simulate a continuum flux error $\sigma(f_\nu)$ determined from the magnitude depth for each photometric band.  We use the continuum fluxes to simulate fluctuations in the HSC photometry due to instrumental noise, in which we convert the magnitude depth for each HSC photometric band into a continuum flux error $\sigma(f_\nu)$.

\begin{figure}[!t]
\begin{center}
\includegraphics[width=0.5\textwidth]{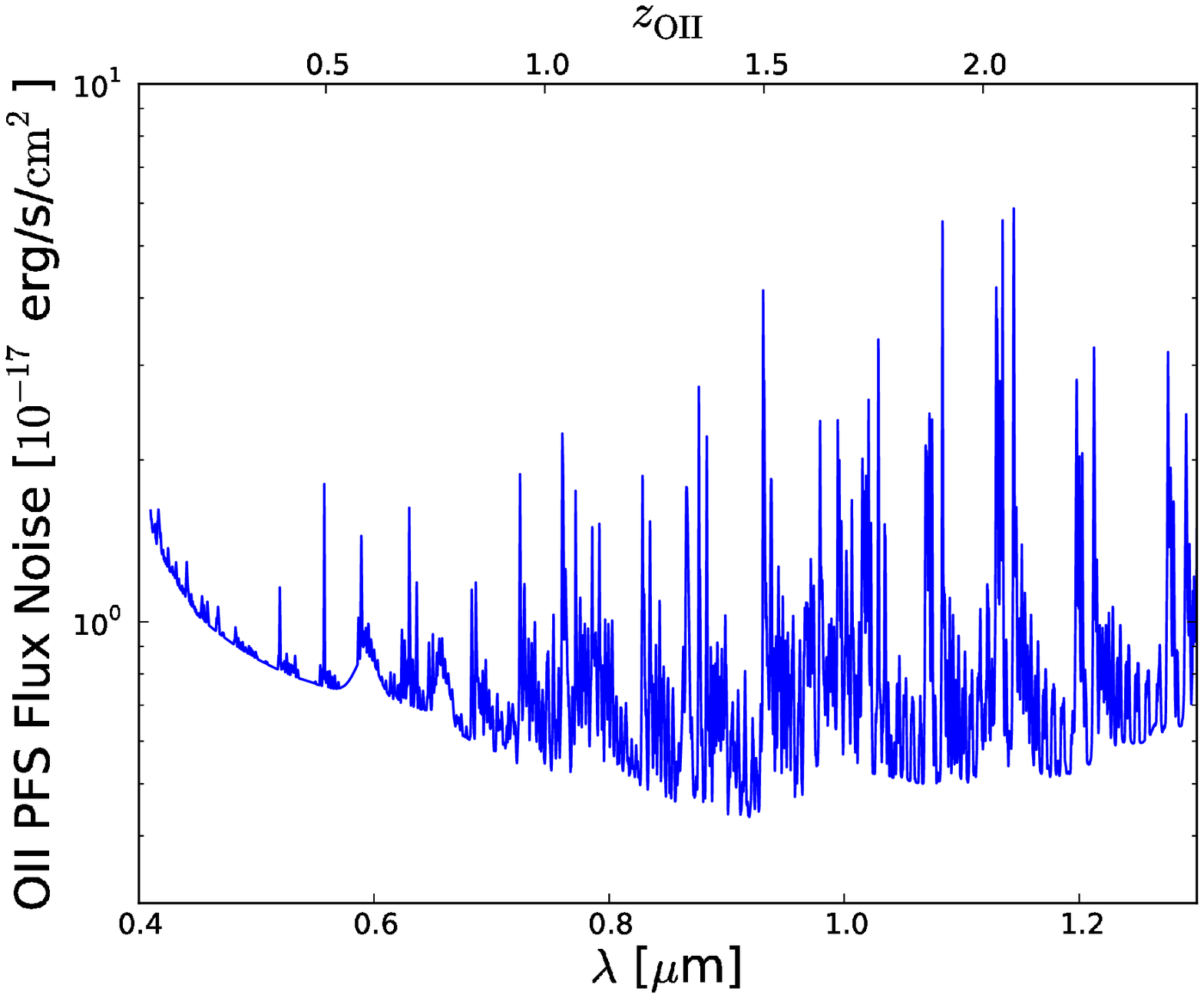}
\includegraphics[width=0.5\textwidth]{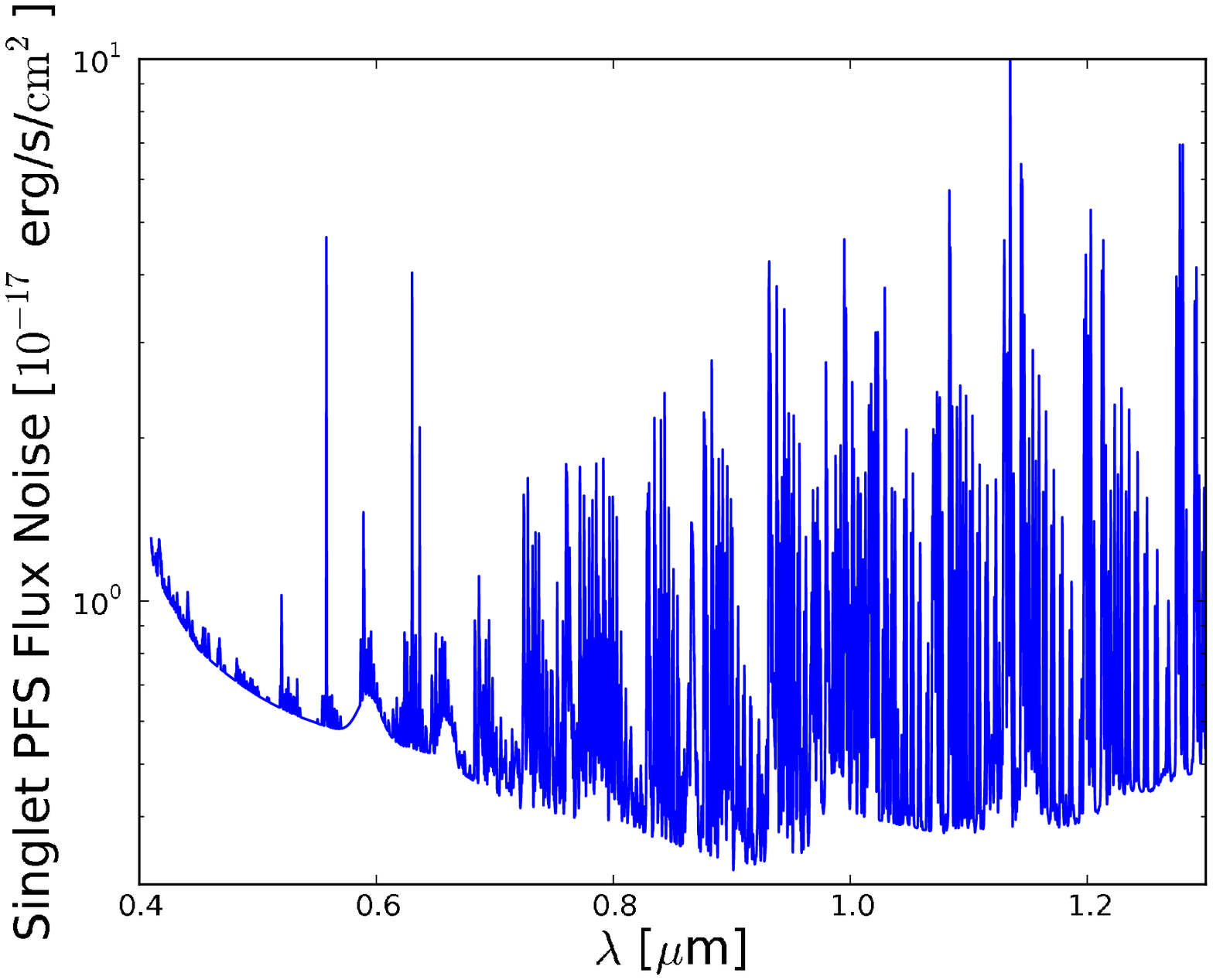}
\caption{\label{F:PFS} The PFS flux noise curve for emission lines.  The upper plot is for the \oii\ doublet and the lower plot is for a single interloper line.}
\end{center}
\end{figure}

We show the doublet and singlet noise curves for PFS in Fig.~\ref{F:PFS}.  We treat \oii\ as a doublet line emission, with the signal consisting of the sum of both lines in the doublet.  The sky lines in the noise curve for the doublet are lower than those for the singlet noise curve because an individual sky line can only disrupt one member of the doublet, making it likely that the other member could still be detectable.\footnote{Note that we neglect increased \oii\ redshift errors caused by disrupted sky lines and blending of the doublet.}  A singlet, however, can be disrupted by a single sky line.  We assume each galaxy has an exponential profile with a half-light radius $r_{\rm eff}=0.3$ arcsec for our noise curve, and we neglect any variation of the source $r_{\rm eff}$ in the PFS analysis.  Note that we treat each of the interlopers that are doublets as separate single emission lines since either line could interlope \oii.  We also require in our analysis that any primary line in a secondary line identification, whether an \oii\ doublet candidate or an interloper, have a $SNR>8.5$, while the secondary line must have $SNR>4$.  We also simulate flux errors due to the PFS detector noise to account for lines with fluxes close to the cutoff that shift above or below it.

The HSC photometry used for target selection can also identify catastrophic redshift errors from interlopers.  We use the CMC to construct the colors $g-r$, $r-i$, and $i-z$ for the objects that pass the PFS target selection and use them to find photometric cuts for redshifts with high interloper fractions.  Note that these photometric cuts are affected by the fluctuations in the HSC photometry introduced earlier.

\subsection{WFIRST mock survey}

WFIRST will have a spectral range of $1.35{\rm \mu m}<\lambda<1.95{\rm \mu m}$ for its cosmology survey, allowing it to map H$\alpha$ ELGs over the redshift range $1.05<z<2$ and \oiii\ (5007\AA) ELGs over the redshift range $1.7<z<2.9$.  Unlike PFS, this mission is space-based, so sky lines will not contaminate the signal.  While its low noise curve will allow the identification of millions of galaxies at high redshifts, interlopers will also be more readily detected.  Also, the WFIRST spectrograph, or grism, will be slitless, removing any requirements for target photometry.

\begin{figure}[!t]
\begin{center}
\includegraphics[width=0.5\textwidth]{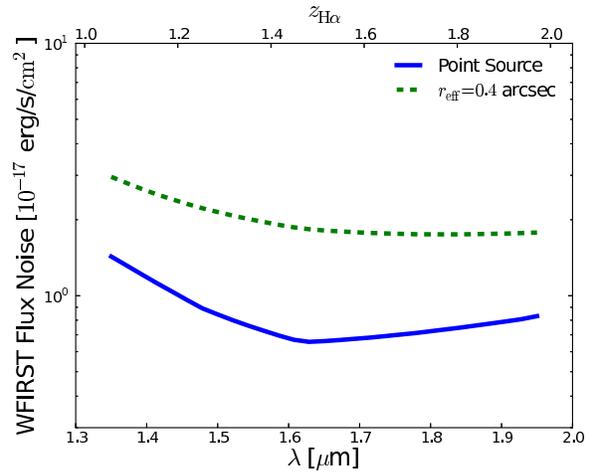}
\caption{\label{F:wfirst} The WFIRST flux noise curves for emission lines.  We plot the cases for a point source (solid) as well as for an effective radius of $r_{\rm eff}=0.4$ arcsec (dashed).}
\end{center}
\end{figure}

The interlopers we consider for WFIRST are the same as for PFS with \oii\ being an additional interloper.  Note that the WFIRST grism's spectral resolution will not be high enough to resolve H$\alpha$ and \nii.  Thus, for both SELGs and interlopers in the WFIRST mock survey, we will combine the H$\alpha$ flux with the \nii\ flux.  Since the WFIRST forecasts in \citet{2013arXiv1305.5422S} assumed a constant ratio $F_{\rm N\,II}/F_{\rm H\alpha}=0.4$, we will attempt to match their forecasts by not using the NII fluxes in the CMC but instead setting $F_{\rm NII}=0.4F_{\rm H\alpha}$.  We also consider the Paschen lines Pa$\alpha$ (1.88 $\mu$m) and Pa$\beta$ (1.28 $\mu$m), which appear in the infrared.  The Paschen lines are not simulated in the CMC, so we determine their (unextincted) fluxes by scaling them with the H$\alpha$ line fluxes, using the atomic line ratios Pa$\alpha$/H$\alpha$=0.119 and Pa$\beta$/H$\alpha$=0.0570 from Appendix B of \citet{2003adu..book.....D}.\footnote{We use the Pa-H$\alpha$ ratios from the low-density limit for $T_e=10^4$K, a typical value, but the temperature dependence is very shallow.}  We show noise curves for WFIRST in Fig.~\ref{F:wfirst}.  In the WFIRST analysis we take into account variations in the noise curve due to $r_{\rm eff}$.  For the WFIRST mock survey, we require a $SNR> 7$ to detect a primary line and $SNR > 4$ to detect a secondary line.  As in the PFS mock survey, we also simulate flux errors due to the instrument.

We also test the use of photometry to remove interlopers from the WFIRST survey.  WFIRST has four photometric bands: F106 (1.06 $\mu$m), F129 (1.29 $\mu$m), F158 (1.58 $\mu$m), and F184 (1.84 $\mu$m).  While these bands are not used for determining targets for spectroscopy, they can still be used to determine locations in color space which can correlate with redshift.  The first three bands correspond to the three near-infrared bands $Y$, $J$, and $H$.  We also consider using photometry from LSST, which has the optical bands $g$, $r$, $i$, $z$, and $y$, where the $y$ band has a similar wavelength range as WFIRST's Y band, though the Y band has a better response on the red end.  We use the CMC to construct all the various color combinations and continuum fluxes $f_\nu$ for all the galaxies.\footnote{LSST, \emph{Euclid}, and WFIRST have slightly different wavelength ranges for each of its similar photometric bands.  We did not color-correct for these differences, but we expect the implications of this approximation to be minor.}  We perform this exercise to determine photometric cuts within redshift bins of $\Delta z=0.2$ that can identify interlopers within the H$\alpha$ and \oiii\ surveys.  As for the PFS mock survey, we use the continuum fluxes to simulate fluctuations in the WFIRST and LSST photometry due to instrumental noise.  We do not include photometry using the F184 band because it had not been tabulated in the current version of the CMC.

\section{PFS results} \label{S:pfsresults}

In this section we present results of our interloper contamination study for the PFS mock survey.  We identify H$\alpha$ as a secondary emission line for identifying OII, and we determine effective secondary emission lines for potential interlopers.  We predict that ELG interloper rates can be reduced to less than 0.2\% using secondary line identification and photometry.

\subsection{Secondary line identification for \oii}

\begin{figure}[!t]
\begin{center}
\includegraphics[width=0.5\textwidth]{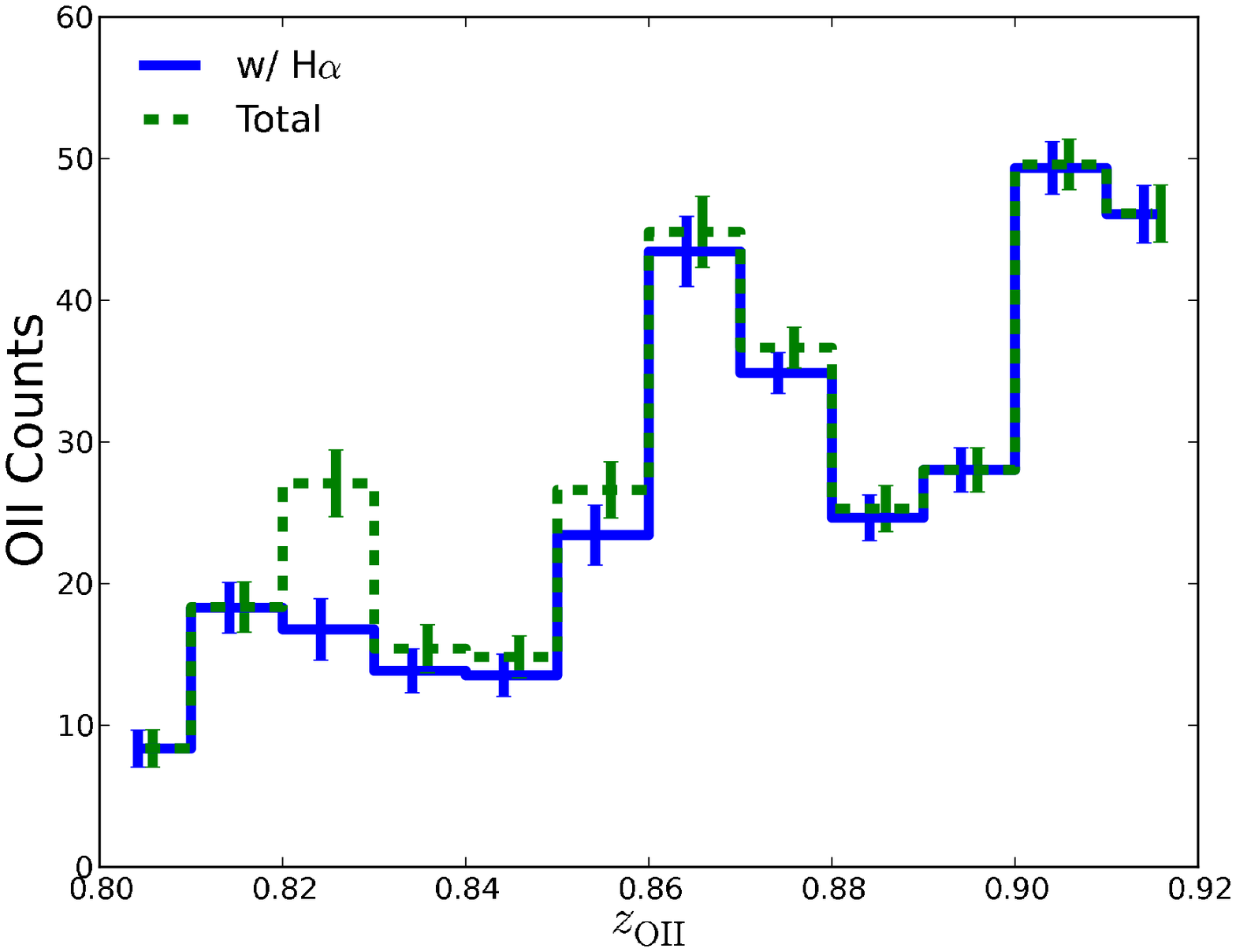}
\caption{\label{F:O2Ha} A histogram of \oii\ emitters detectable by PFS in redshift bins $\Delta z=0.01$.  The dashed curve is the histogram for all \oii\ emitters with $SNR>8.5$, and the solid curve is the histogram for those with a detectable H$\alpha$ line.  The error bars denote variations in the distribution of detected \oii\ counts, are determined using simulated flux noise, and are slightly shifted for clarity.  The decrement in the fraction of \oii\ emitters with H$\alpha$ lines at $z_{\rm OII}=0.825$ is due to H$\alpha$ encountering a sky line at $\lambda_{\rm Ha}\simeq1.2\mu$m.}
\end{center}
\end{figure}

We tested several emission lines to determine any that could serve as a secondary emission line for \oii, and we found that the best choice by far was the H$\alpha$ line, mainly because it is a much stronger emission line than any other lines within PFS's wavelength range.  Since the PFS wavelength window ends at 1.26$\mu$m, H$\alpha$--\oii\ line identification can only be used for $z_{\rm OII}<0.92$.  Over the range $0.8<z_{\rm OII}<0.92$, we find that 94\% of \oii\ emitters in our PFS mock have a detectable H$\alpha$ line.  As a function of redshift, we see in Fig.~\ref{F:O2Ha} that in most redshift bins in this redshift range, most \oii\ lines in the PFS mock are identified using the H$\alpha$ line.  There is a large decrement in the fraction of \oii\ emitters with H$\alpha$ lines at $z_{\rm OII}=0.825$, due to H$\alpha$ encountering a sky line at $\lambda_{\rm Ha}\simeq1.2\mu$m, which causes the percentage of \oii\ emitters paired with H$\alpha$ to drop to 60\%.

\subsection{Secondary line identification for interlopers} \label{S:pfs2line}

\begin{figure*}
\begin{center}
\includegraphics[width=\textwidth]{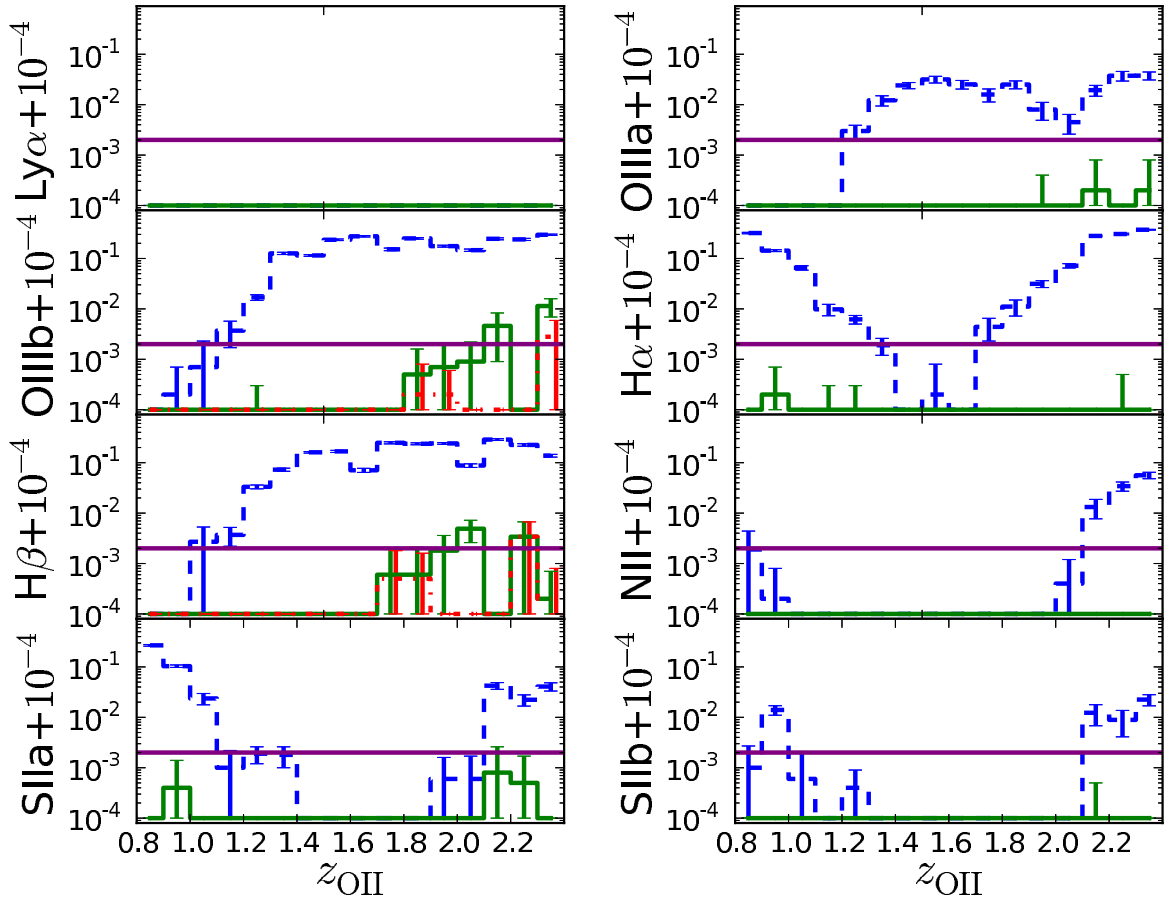}
\caption{\label{F:pairs} A histogram of $f({\rm OII}-X,z_{\rm OII})$ in redshift bins $\Delta z=0.1$ for the PFS survey.  The dashed, blue curve is the histogram for all interlopers with $SNR>8.5$, and the solid, green curve is the histogram for those that cannot be ruled out as an OII candidate using secondary line identification.  The dash-dotted, red line in the \oiii\ 5007 (H$\beta$) plot is the histogram including SLI with the \oiii\ 4959 line (photometry).  The error bars denote variations in the distribution of interloper fractions determined using simulated flux noise.  The solid, purple line denotes the target interloper fraction $f=0.2$\%.  Note that the interlopers that are members of doublets are labeled $a$ and $b$ which denote the shorter and longer wavelength line, respectively.}
\end{center}
\end{figure*}

We repeat the exercise from the previous section for potential interlopers of the \oii\ doublet to see if they could be identified.  For Ly$\alpha$, the PFS photometric cuts alone eliminate them from the sample.  For the other interlopers, we determine other lines that could help rule them out as \oii\ candidates.  We list the interlopers with their secondary lines in Table \ref{T:pairs}.  In Fig.~\ref{F:pairs}, we plot $f({\rm OII}-{\rm Int},z_{\rm OII})$ for all the interlopers.  Some of the lines, like \oiii, can have increased interloper rates because they tend to have fluxes just below the detection threshold, which then shift due to flux noise. The flux noise tends to cause more of them to shift above the threshold than to shift below the threshold.  We also see that secondary line identification can eliminate H$\alpha$, \sii, and \nii\ entirely.  This method also decreases rates for \oiii\ 4959/5007 and H$\beta$ to less than 1\% contamination \emph{over each small redshift bin}.  Of course, the CMC is not perfect, and the luminosity functions on which this catalog is based, including for H$\alpha$ and \oii, are uncertain.  Overall, it seems that the remaining interlopers that may contaminate the PFS results are \oiii\ and H$\beta$ at \oii\ redshifts $z_{\rm OII} > 2$.


We reduce the \oiii\ emission lines further by using secondary line identification for the doublet with a SNR cutoff of $Q_c=1$ (see Sec.~\ref{S:2line}).  We also include the effect from \oii\ lines being identified as \oiii\ interlopers due to a statistical fluctuation appearing as an \oiii\ doublet partner.  We model the contribution of fluctuations from the continuum by estimating according to our formalism that up to $P_2=16\%$ of all \oii\ lines will appear to be \oiii\ lines and be cut due to our low cutoff for the \oiii\ 4959 line.  We still find, however, that this secondary line identification scheme allows us to identify \oiii\ 5007 interlopers in our mock PFS survey.  Specifically, the resulting \oiii\ 5007 interloper fraction is less than 0.2\%, except in the highest redshift bin ($z_{\rm OII}=2.35$) where the fraction is slightly above 0.2\%. Thus, we conclude that \oiii\ interlopers should be removable for PFS using secondary line identification.

We also consider how our PFS results are affected if any PFS target selection cuts (see Eq.~\ref{E:pfstargets} were removed.  We find that if only the $22.8\leq g\leq 24.2$ cut is kept, the interloper fractions increase well past our 0.2\% target.  This is the case particularly for \oiii\ interlopers, which exhibit a contamination level of $f=10$\% in this scenario.
\begin{table*}
\begin{center}
\caption{\label{T:pairs} Potential interlopers with the secondary emission lines used to identify them.}
\vspace{0.5cm}
\begin{tabular}{c|ccc}
\hline
Line&Secondary (PFS \oii\ )&Secondary (WFIRST H$\alpha$)&Secondary (WFIRST \oiii\ )\\
\hline
Ly$\alpha$&N/A&N/A&N/A\\
\oii&N/A&\oiii,H$\beta$&\oiii,H$\beta$\\
H$\beta$&H$\alpha$,\oii&\oiii&\oiii,\oii\\
\oiii&\oii&\oii,H$\alpha$,H$\beta$&N/A\\
H$\alpha$&\oii,\oiii&N/A&\oiii,\sii\\
\nii&\oii,\oiii&N/A&\oiii,\sii\\
\sii&H$\alpha$&H$\alpha$&H$\alpha$,\oiii\\
\hline
\end{tabular}\end{center}
\end{table*}

\subsection{Photometry}

H$\beta$ is the remaining interloper in the PFS survey with an interloper fraction $f(z_{\rm OII}\sim2)\simeq0.6$\%, and we attempt to use HSC photometry to remove these interlopers.  For H$\beta$ interlopers in the range $z_{\rm OII}=$ 1.9--2.1 ($z_{\rm H\beta}=$ 1.22--1.38), we find these interlopers fail the photometric cut $i-z<0.67(r-i)+0.167$.  Based on the CMC, this cut removes only $\sim2$\% of \oii\ emitters, while not identifying only $\sim0.1$\% of H$\beta$ interlopers.  Using this cut eliminates  the remaining H$\beta$ interlopers in this redshift range.  We also see an interloper rate of 0.3\% in the range $z_{\rm OII}=$ 2.2--2.3 ($z_{\rm H\beta}=$ 1.45--1.53); however, we were unable to find a satisfactory photometric cut in this redshift range using HSC bands.

\begin{figure}
\begin{center}
\includegraphics[width=0.5\textwidth]{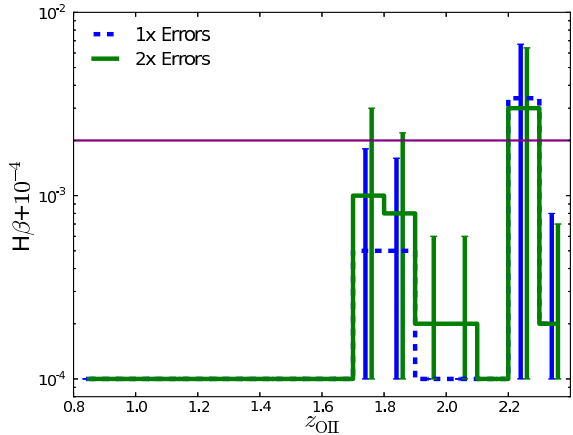}
\caption{\label{F:hbdiff} A histogram of $f({\rm OII}-H\beta,z_{\rm OII})$ in redshift bins $\Delta z=0.1$ for the PFS survey.  The dashed, blue curve includes all secondary line identification and photometric cuts (represented by the dash-dotted line in Fig.~\ref{F:pairs}) while the solid, green line is the same with double the photometric errors.  The solid, purple line denotes the target interloper fraction $f=0.2$\%.  We see that the increase in interlopers is not enough to cause further significant contamination.}
\end{center}
\end{figure}

It is also likely that the Balmer + 4000 \AA\ break feature is weaker in the real galaxies than in the CMC (see \S\ref{S:mock}). To test our sensitivity to this effect, we repeat our analysis of the H$\beta$ photometric cut while doubling the errors in all the photometric bands (this is a crude proxy for halving the break stength). The result is shown in Fig.~\ref{F:hbdiff}, where we see that the number of interlopers does increase, but the contamination level is still less than $0.2$\% at most redshifts.

\section{WFIRST results} \label{S:wfresults}
In this section we determine the interloper rates for WFIRST's H$\alpha$ and \oiii\ (5007\AA) surveys.  It does appear that WFIRST could potentially exhibit greater interloper contamination than PFS.  After implementing secondary line identification, the main interlopers for the surveys are \oii, \oiii\ (5007\AA), H$\alpha$, and the Paschen lines.  We do find that the photometric bands from LSST and WFIRST can reduce the remaining interlopers in the H$\alpha$ survey to less than 0.2\% at all redshifts.  However, the \oiii\ survey still exhibits interloper rates of up to a few percent even after photometric cuts. The fundamental reason for this appears to be that at WFIRST flux levels, there are many more H$\alpha$ emitters than \oiii\ emitters, hence the contamination of the \oiii\ sample by H$\alpha$ can be significant even if secondary line identification and photometric cuts exclude most of the H$\alpha$ contaminants.

\subsection{H$\alpha$ survey}
\begin{figure}[!t]
\begin{center}
\includegraphics[width=0.5\textwidth]{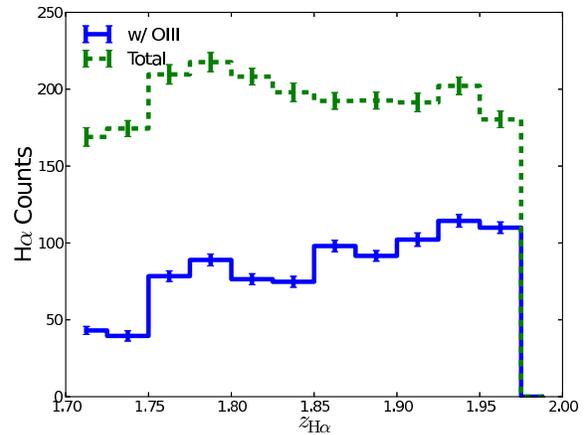}
\caption{\label{F:HaO3} A histogram of H$\alpha$ emitters detectable by WFIRST in redshift bins $\Delta z=0.025$.  The dashed curve is the histogram for all H$\alpha$ emitters with $SNR>7$, and the solid curve is the histogram for those with a visible \oiii\ (5007\AA) line.  The error bars denote variations in the distribution of detected H$\alpha$ counts determined using simulated flux noise.}
\end{center}
\end{figure}

Our attempt to find a candidate for the secondary line identification of H$\alpha$, whose redshift range for the WFIRST survey is $1<z_{\rm H\alpha}<2$, was unsuccessful. \oiii\ (5007\AA) is the best candidate in that it appears with H$\alpha$ over the longest spectral range, specifically over $z_{\rm H\alpha}>1.7$, and \oiii\ is one of the brightest emission lines. However, only 43\% of H$\alpha$ lines with $z_{\rm H\alpha}>1.7$ detected by WFIRST will have a corresponding \oiii\ line with $SNR>4$ within the WFIRST spectral range. We plot in Fig.~\ref{F:HaO3} a comparison of number counts for detected H$\alpha$ lines with and without secondary line identification.  Thus, we will not attempt to use secondary line identification to confirm H$\alpha$ lines, but instead assume all single lines are H$\alpha$ unless we can prove they are interlopers.


\begin{figure*}
\begin{center}
\includegraphics[width=\textwidth]{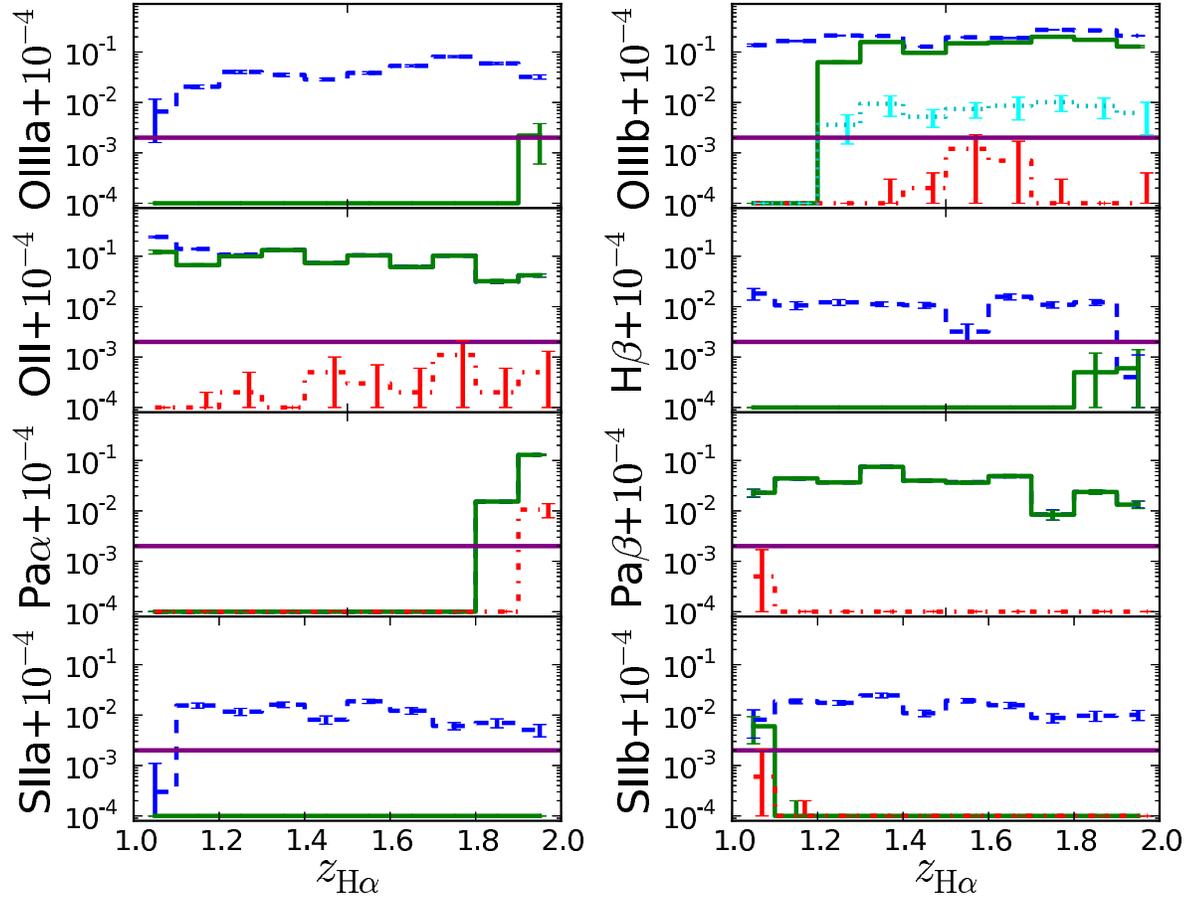}
\caption{\label{F:halines} A histogram of $f({\rm H\alpha}-X,z_{\rm H\alpha})$ in redshift bins $\Delta z=0.1$ for the WFIRST H$\alpha$ survey.  The dashed, blue curve is the histogram for all interlopers with $SNR>7$, and the solid, green curve is the histogram for those that cannot be ruled out as an H$\alpha$ candidate using secondary line identification.  The dotted, cyan line in the \oiii\ 5007 plot is the interloper rate after implementing secondary line identification using the \oiii\ 4959 line.  The dash-dotted, red line includes photometric cuts described in Sec.~\ref{S:wphoto}.  The error bars denote variations in the distribution of interloper fractions determined using simulated flux noise.  The solid, purple line denotes the target interloper fraction $f=0.2$\%.  Note that the interlopers that are members of doublets are labeled $a$ and $b$ which denote the shorter and longer wavelength line, respectively.}
\end{center}
\end{figure*}

We also find emission lines that serve as efficient secondary lines for potential interlopers, which are listed in Table \ref{T:pairs}.  As with the PFS survey, Ly$\alpha$ is below the flux cut for WFIRST.  The interloper rates for the remaining lines are shown in Fig.~\ref{F:halines}.  We see that although secondary line identification eliminates most of the interloper fractions down to $<1$\%, \oiii\ and \oii\ appear to remain with high interloper rates of around 30\%, although we are able to reduce \oiii\ down to 10\% by identifying \oiii\ 5007 using the \oiii\ 4959 line (see Sec.~\ref{S:pfs2line}.  The limitation in eliminating these interlopers is partially due to the limited spectral range of WFIRST, which limits the use of H$\alpha$ to identify \oiii\ and \oiii\ to identify \oii.  The Paschen lines Pa$\alpha$ and Pa$\beta$ have no viable candidates for secondary line identification because their wavelengths are too far apart from other lines and each other.  

\subsection{OIII survey}
\begin{figure}[!t]
\begin{center}
\includegraphics[width=0.5\textwidth]{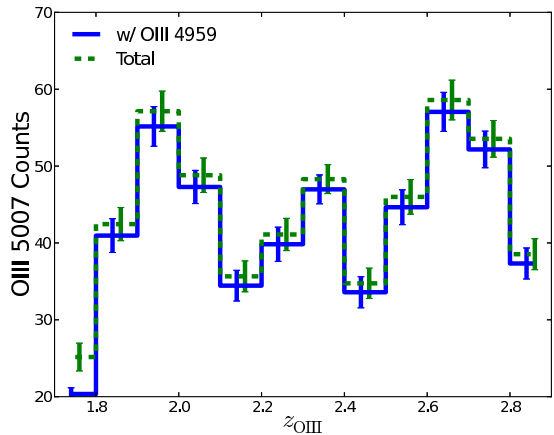}
\caption{\label{F:O3O3} A histogram of \oiii\ (5007\AA) emitters detectable by WFIRST in redshift bins $\Delta z=0.025$.  The dashed curve is the histogram for all \oiii\ emitters with $SNR>7$, and the solid curve is the histogram for those with an \oiii\ 4959 line with $SNR>1$.  The error bars denote variations in the distribution of detected \oiii\ counts determined using simulated flux noise.}
\end{center}
\end{figure}

The WFIRST \oiii\ (5007\AA) survey, which has a redshift range of $1.7<z_{\rm OIII}<2.9$, can use the \oiii\ 4959 line ($SNR>1$) for secondary line identification.  As seen in Fig.~\ref{F:O3O3}, using the \oiii\ 4959 line keeps almost all (96\%) of the \oiii\ 5007 sources.

\begin{figure*}
\begin{center}
\includegraphics[width=\textwidth]{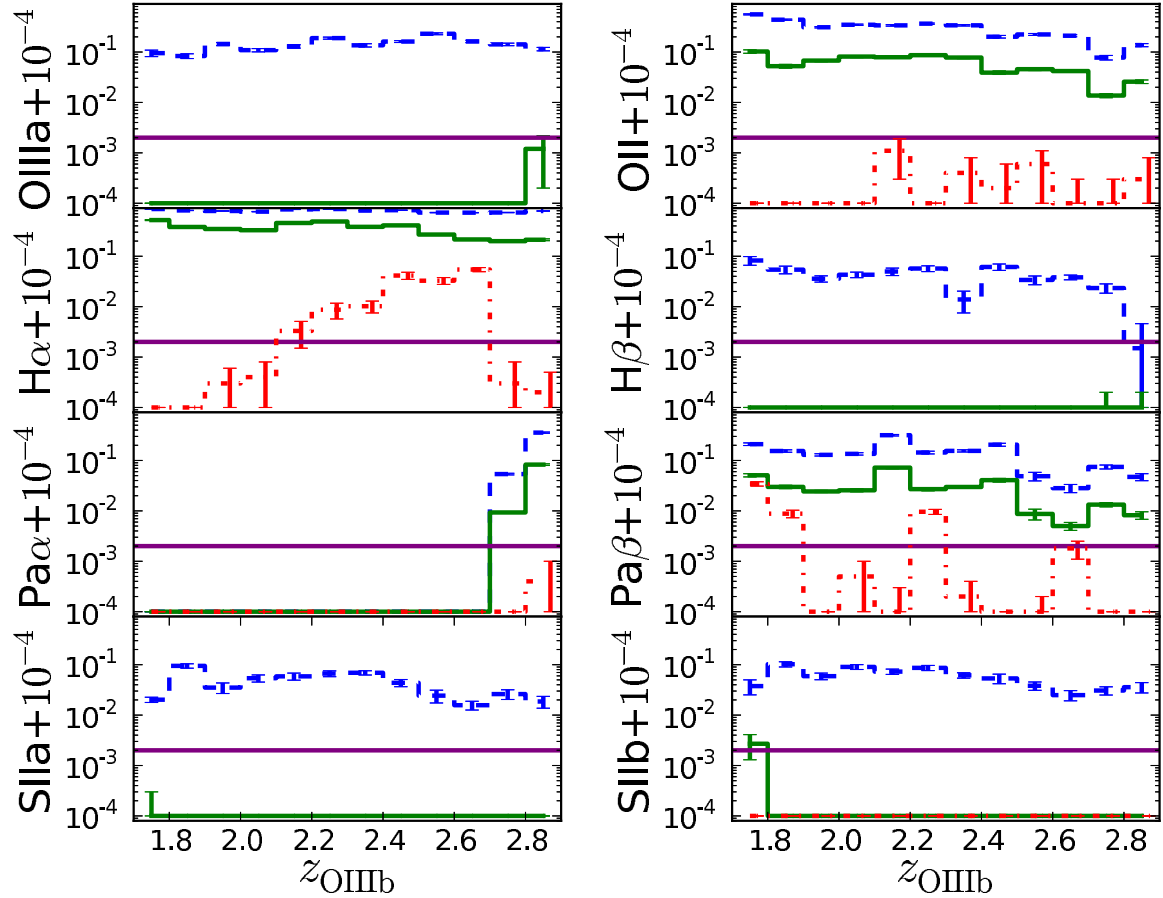}
\caption{\label{F:o3lines} A histogram of $f({\rm OIII}-X,z_{\rm OIII})$ in redshift bins $\Delta z=0.1$ for the WFIRST OIII survey.  The format of the figure is similar to Fig.~\ref{F:halines}.}
\end{center}
\end{figure*}

Using the \oiii\ 4959 line to identify \oiii\ 5007 will not eliminate all interlopers perfectly; random fluctuations from the continuum at the right wavelength can cause an interloper to appear to have a corresponding \oiii\ 4959 line.  As before (see Sec.~\ref{S:2line}), the CMC does not simulate fluctuations from the continuum, so we have to estimate how often a false \oiii\ 4959 line will appear.  Based on our formalism, we expect that only $P_2=16\%$ of interlopers will have a corresponding \oiii\ 4959 line for $Q_c=1$ due to statistical fluctuations, so we multiply the interloper number counts in our analysis by $P_2$.  After testing potential interlopers for secondary line candidates, we find that \oii, H$\alpha$, and the Paschen lines are the major interlopers left after secondary line identification.  The interlopers' secondary lines are listed in Table \ref{T:pairs}, and the interloper rates are shown in Fig.~\ref{F:o3lines}.  Pa$\alpha$ does not become a problem until $z_{\rm OIII}>2.7$, so we may be able to eliminate this interloper by using secondary line identification for \oiii\ 5007 using \oiii\ 4959.  The other three interlopers appear to have high interloper rates, particularly at the lower redshifts for H$\alpha$ and Pa$\beta$ and higher redshifts for \oii.

For WFIRST, we see that H$\alpha$ and \oiii\ could heavily contaminate each others' surveys even after secondary line identification, and \oii\, Pa$\alpha$, and Pa$\beta$ could contaminate both surveys.  It appears that better methods will be required to reduce these interlopers.

\subsection{Photometry} \label{S:wphoto}

We attempt to use photometry to eliminate more interlopers from the WFIRST surveys.  Since LSST should be online long before WFIRST, we explore the use of both WFIRST and LSST photometric bands to remove the interlopers remaining after secondary line identification.  In Tables \ref{T:photoha} and \ref{T:photoo3}, we list the best photometric cuts to identify \oiii, \oii, \sii, Pa$\alpha$, and Pa$\beta$ interlopers in the H$\alpha$ survey and H$\alpha$, \oii, \sii, Pa$\alpha$, and Pa$\beta$ interlopers in the \oiii\ survey.  We also consider, based on the CMC, how often true SELGs fail the cuts and interlopers pass the cuts.  The error rates are all less than 5\%, with most less than 1\%.

\begin{figure*}
\begin{center}
\includegraphics[width=\textwidth]{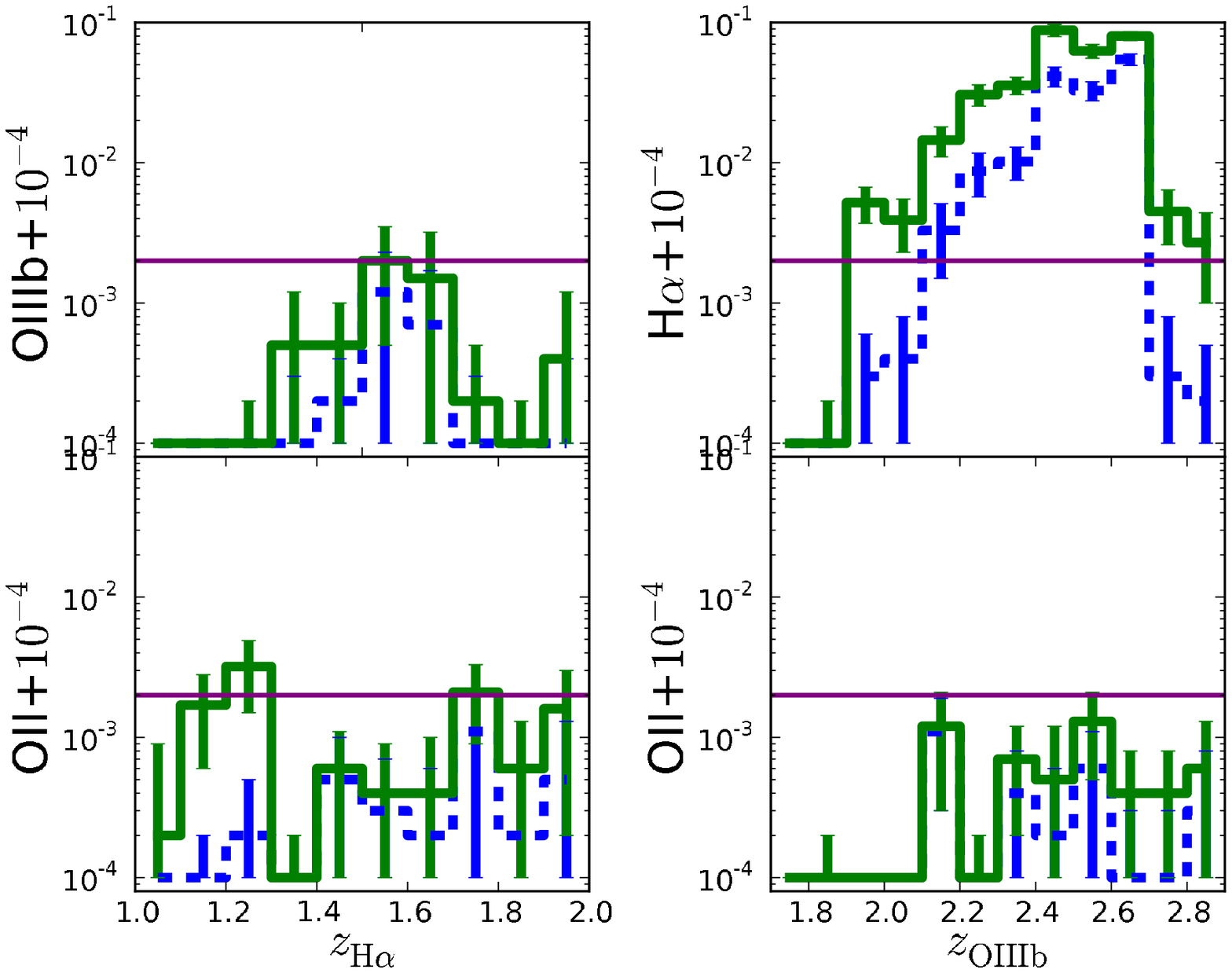}
\caption{\label{F:wfdiff} Histograms of interloper fractions for the WFIRST H$\alpha$ and \oiii\ surveys.  In each plot, the dashed, blue curve includes all secondary line identification and photometric cuts (represented by the dash-dotted lines in Figs.~\ref{F:halines} and \ref{F:o3lines}) while the solid, green line is the same with double the photometric errors.  The solid, purple line denotes the target interloper fraction $f=0.2$\%.  Also the left column is for the H$\alpha$ survey, while the right column is for the \oiii\ survey.  We see that these interloper rates increase significantly when the photometric errors increase.}
\end{center}
\end{figure*}

When these photometric cuts are implemented for interlopers in the H$\alpha$ survey, all interloper rates decrease to less than 0.2\%, satisfying our requirement, except for Pa$\alpha$, which still has a 1\% interloper fraction in the highest redshift bin.  For the \oiii\ survey, we find that H$\alpha$ and Pa$\beta$ are still significant interlopers, with interloper rates due to H$\alpha$ greater than 1\% at some redshifts.  In the same manner as for PFS, we consider the effect of doubling the photometric errors on these cuts, in order to consider the effects of breaks that are weaker than those in the CMC. The interloper rates for \oii\ and \oiii\ in the H$\alpha$ survey and for \oii\ and H$\alpha$ in the \oiii\ survey increase significantly, as we see in Fig.~\ref{F:wfdiff}, with \oii\ in the H$\alpha$ survey passing the 0.2\% level and H$\alpha$ in the \oiii\ survey even reaching 10\% contamination at some redshifts.

Part of the reason H$\alpha$ is such a large contaminant is that H$\alpha$-emitters are the dominant population in the WFIRST survey, as by design, so a very large number of H$\alpha$ interlopers need to be eliminated to reduce the contamination level to less than 0.2\%.  Also, it can be shown that for $2.3<z_{\rm OIII}<2.7$, the \oiii\ color locus for the relevant photometric cut is close enough to the H$\alpha$ interloper color locus that random fluctuations cause them to overlap, greatly reducing the power of the cut.  On the extreme ends of the \oiii\ survey redshift range, the color loci are farther apart, which reduces the population of H$\alpha$ interlopers considerably.  H$\alpha$ interlopers could be mitigated somewhat by identifying them in the H$\alpha$ survey, but this would inevitably cause some \oiii\ lines misidentified as H$\alpha$ lines to be removed.  For these cases, a deep spectroscopic training samples may be required to accurately measure the interloper fraction.

\subsection{Y-band Photometry}

Most of the photometric cuts needed to identify interlopers require the $Y$ band, a fairly new photometric band for LSS surveys. To motivate the required $Y$-band depth for these surveys, we determine the distribution of $Y$-band magnitudes expected in the WFIRST survey, both for the SELGs and the major contaminants that are not removed through secondary line identification. In our analysis we simulate fluctuations in the WFIRST $Y$-band magnitude by converting the magnitude depth of the band ($Y=26.7$ for point sources, 5$\sigma$) into a continuum flux error.  The fluctuations are then added to the continuum $Y$-band fluxes of the emitters and converted back to apparent magnitudes.

\begin{figure}
\begin{center}
\includegraphics[width=0.5\textwidth]{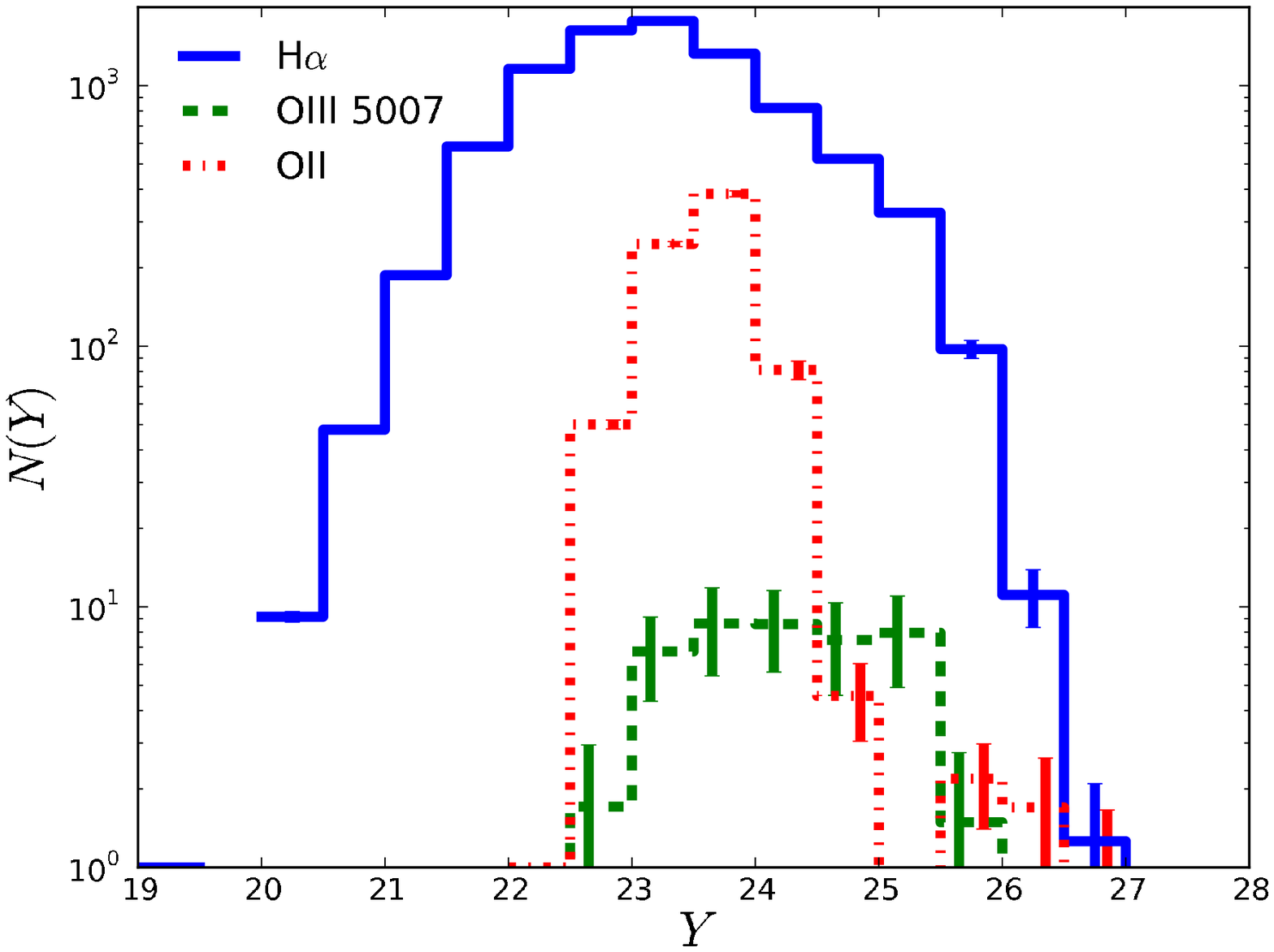}
\includegraphics[width=0.5\textwidth]{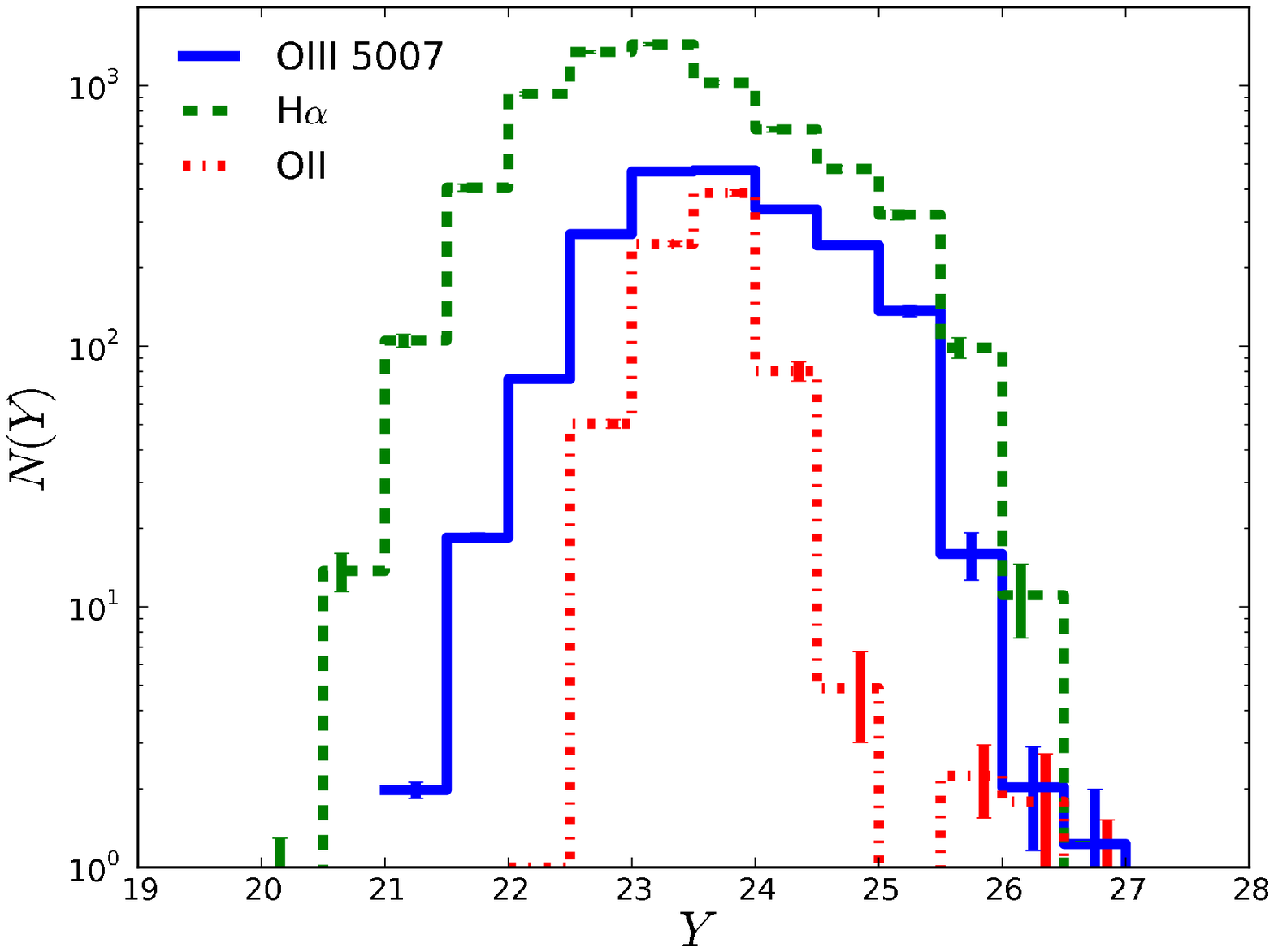}
\caption{\label{F:yHaO3} A histogram of the distribution of $Y$-band magnitudes for the WFIRST H$\alpha$ survey (top) and the OIII survey (bottom) plotted in $Y$-magnitude bins $\Delta Y=0.5$.  Each panel includes the survey emission line galaxies along with contaminants that cannot be identified using secondary line identification, requiring photometry to be removed. The error bars denote variations in the distribution of $Y$-band magnitudes determined using simulated flux noise in the lines and the $Y$ magnitude.  We see from these plots that $Y$-band photometry must be deeper than 26th magnitude in order to measure colors for ``almost all'' of the sample.}
\end{center}
\end{figure}

We plot the $Y$-band magnitude distributions in Fig.~\ref{F:yHaO3}.  It can be seen that all of the galaxies and interlopers will have a $Y$-band magnitude $Y<26.7$.  While a deeper $Y$ survey would be preferable, WFIRST in its current configuration should be able to determine the $Y$ magnitude for most of the galaxies (SELGs and interlopers) in the survey.  Note that the LSST ($y<25$) or Euclid ($Y<24$) photometry in this band would miss a significant number of the source galaxies, and so at present it appears there is a need for the WFIRST $Y$-band photometry in order to carry out the galaxy redshift survey program. However, the WFIRST survey will require LSST optical bands to implement many of the other photometric cuts needed to remove interlopers.

\begin{table*}
\begin{center}
\caption{\label{T:photoha} Photometric cuts for WFIRST H$\alpha$ survey as a function of redshift.}
\vspace{0.5cm}
\begin{tabular}{c|cc}
\hline
$z_{\rm H\alpha}$&OIII 5007 photometric cut&OII photometric cut\\
\hline
0.9--1.1&$Y-J<1.02(i-z)+0.11$&$r-i>0.52(g-r)+0.052$\\ 
1.1--1.3&$Y-J<1.1(i-z)+0.15$&$Y-J<1.4(i-z)-0.2$\\
1.3--1.5&$Y-J<1.935(z-Y)+0.161$&$\left.\begin{array}{l}i-z>0.36(g-r)-0.02\\\&\,r-i >0.56(g-r)-0.12\end{array}\right.$\\
1.5--1.7&$J-H<\left\{\begin{array}{ll}4.2(z-Y)-0.1&\mbox{if $z-Y<0.2$};\\0.6(z-Y)+0.7&\mbox{if $z-Y>0.2$}\end{array}\right.$&$r-i>0.68(g-r)-0.364$\\
1.7--1.9&$J-H<0.714(Y-J)+0.071$&$r-i>0.62(g-r)-0.438$\\
1.9--2.0&$J-H<Y-J+0.1$&$r-i>0.75(g-r)-0.75$\\
\hline
&SII 6731 photometric cut ($z_{\rm H\alpha}=0.9-1.1$)&Pa$\alpha$ photometric cut ($z_{\rm H\alpha}=1.8-2.0$)\\
\hline
&$J-H > 0.91(Y-J)-0.228$&$Y-J>2.33(z-Y)+0.05$\\
\hline
&Pa$\beta$ photometric cut&\\
\hline
0.9--1.1&$Y-J<0.89(i-z)$&\\
1.1--1.3&$i-z>0.583(g-r)$&\\
1.3--1.5&$i-z>0.3(g-r)+0.15$&\\
1.5--1.7&$z-Y>0.167(r-i)+0.033$&\\
1.7--1.9&$Y-J>2.0(i-z)-0.1$&\\
1.9--2.0&$Y-J>1.9(i-z)$&\\
\hline
\end{tabular}\end{center}
\end{table*}


\begin{table*}
\begin{center}
\caption{\label{T:photoo3} Photometric cuts for WFIRST OIII survey as a function of redshift.}
\vspace{0.5cm}
\begin{tabular}{c|cc}
\hline
$z_{\rm OIII}$&H$\alpha$ photometric cut&OII photometric cut\\
\hline
1.7--1.9&$Y-J>1.5(i-z)$&$J-H<0.94(Y-J)+0.04$\\ 
1.9--2.1&$Y-J>2.35(z-Y)+0.2175$&$J-H<1.11(Y-J)-0.1$\\
2.1--2.3&$Y-J>3.0(z-Y)+0.1$&$r-i>0.58(g-r)-0.16$\\
2.3--2.5&$J-H>6.06(z-Y)-0.621$&$i-z>0.4(g-r)-0.2$\\
2.5--2.7&$J-H>3.5(z-Y)$&$i-z>0.42(g-r)-0.368$\\
2.7--2.9&$J-H>0.947(Y-J)-0.226$&$i-z>0.4(g-r)-0.4$\\
\hline
&SII 6731 photometric cut ($z_{\rm OIII}=1.7-1.9$)&Pa$\alpha$ photometric cut ($z_{\rm OIII}=2.7-2.9$)\\
\hline
&$Y-J>1.625(i-z)+0.1$&$J-H>0.81(Y-J)+0.16$\\
\hline
&Pa$\beta$ photometric cut&\\
\hline
1.7--1.9&$Y-J>1.5(i-z)+0.2$&\\ 
1.9--2.1&$Y-J>2.556(z-Y)+0.2556$&\\
2.1--2.3&$Y-J>3.25(z-Y)+0.2$&\\
2.3--2.5&$z-Y>0.714(r-i)-0.1572$&\\
2.5--2.7&$z-Y>0.25(i-z)-0.05$&\\
2.7--2.9&$J-H>1.167(z-Y)+0.1167$&\\
\hline
\end{tabular}\end{center}
\end{table*}


\section{Conclusions} \label{S:conc}
We predict the effects of interlopers on cosmological parameter estimates from upcoming and future LSS surveys.  We construct a formalism describing our interloper contamination biases estimates of the power spectrum as well as cosmological parameter estimates.  Interloper fractions $f>0.5\%$ can bias the measured power spectrum from LSS surveys, particularly on smaller scales, while $f>0.2\%$ will significantly increase error estimates.  Also, contamination levels greater than 0.15--0.3\% can significantly bias the growth rate measurements from upcoming surveys.

Using the CMC, we estimate the amount of interloper contamination that can occur in upcoming surveys.  Note again that our analysis with the CMC is idealized in that it does not treat blended objects and is limited by the bank of SED templates in the catalog, but it represents a good ``first look'' at how serious the problem might be, and what mitigation strategies might be employed.  In our estimates we simulate the interloper cleaning methods of secondary line identification and photometric identification.  While we expected secondary line identification to be effective, we find specifically that this method can potentially eliminate most interlopers from the PFS survey, but it will leave large interloper fractions in the future WFIRST survey.  We find photometric cuts are well able to identify interlopers in WFIRST.  We predict that WFIRST can exhibit contamination levels up to 30\% using only secondary line identification, which would be catastrophic for dark energy measurements, but photometric cuts can reduce contamination in WFIRST to less than 0.2\% for most redshifts in the H$\alpha$ survey.  In the \oiii\ survey, large interloper fractions remain after photometric cuts. Deep spectroscopic surveys will be required to measure the significant interloper fractions that remain in the PFS and WFIRST surveys, particularly for H$\alpha$ in the \oiii\ survey if it cannot be readily identified from the H$\alpha$ survey. We caution again that these are point estimates based on the current ``community standard'' mock catalog, and will evolve as the fidelity of the mock catalogs and ultimately the real data from these instruments become available.

These strategies are applicable to many upcoming emission line surveys, including \emph{Euclid}; however, the actual implementation will depend heavily on spectroscopic and photometric sensitivities and wavelength ranges, as well as survey areas which will determine available ancillary photometric surveys.  This analysis was performed specifically for optical/NIR surveys, but the strategies can also be implemented in surveys at other wavelengths, e.g. millimeter wavelengths, where there are multiple lines and where instrumental or atmospheric transmission considerations limit the available spectral range.  In addition, causes of scale-dependent clustering bias such as non-Gaussianity and neutrino masses will have varying sensitivities to the interloper fraction.  Future work should determine acceptable interloper fractions for other cosmological parameters as well as the implementation of these strategies for other upcoming surveys.

\begin{fund}
AP was supported by the McWilliams Fellowship of the Bruce and Astrid McWilliams Center for Cosmology. Part of the research described in this paper was carried out at the Jet Propulsion Laboratory, California Institute of Technology, under a contract with the National Aeronautics and Space Administration. AP was supported by an appointment to the NASA Postdoctoral Program at the Jet Propulsion Laboratory, California Institute of Technology, administered by Oak Ridge Associated Universities through a contract with NASA.  CH is supported by the David and Lucile Packard Foundation, the Simons Foundation, and the U.S. Department of Energy.  AR is supported by the John Templeton Foundation.
\end{fund}

\begin{ack}
We thank P.~Capak, D.~Eisenstein, S.~Ho, M.~Seiffert, and M.~Takada for helpful comments and useful discussions.  We also thank J.-P.~Kneib for his assistance with the COSMOS Mock Catalog, in particular for providing access to the $Y$-band photometry.
\end{ack}


\end{document}